\documentclass[sigconf]{acmart}
\AtBeginDocument{%
  }

\setcopyright{acmlicensed}

\copyrightyear{2026}
\acmYear{2026}
\setcopyright{cc}
\setcctype{by}
\acmConference[CHI '26]{Proceedings of the 2026 CHI Conference on Human Factors in Computing Systems}{April 13--17, 2026}{Barcelona, Spain}
\acmBooktitle{Proceedings of the 2026 CHI Conference on Human Factors in Computing Systems (CHI '26), April 13--17, 2026, Barcelona, Spain}
\acmPrice{}
\acmDOI{10.1145/3772318.3790530}
\acmISBN{979-8-4007-2278-3/2026/04}

\acmSubmissionID{6299}
\newcommand{\ouyang}{\textcolor{black}}

\newcommand{\ouyangre}[1]{\textcolor{black}{#1}}

\usepackage{makecell}

\usepackage{amssymb}  % for \checkmark
\usepackage{multirow}
\usepackage{graphicx}
\usepackage{cleveref} 
\usepackage{listings}
\usepackage{colortbl}
\usepackage{enumitem}

\begin{document}

\title{\textit{CommSense}: Facilitating Bias-Aware and Reflective Navigation of Online Comments for Rational Judgment}


\author{Yang Ouyang}
\orcid{0009-0000-5841-7659}
\authornote{Both authors contributed equally to this research.}
\affiliation{%
  \institution{School of Information Science and Technology \\ ShanghaiTech University}
  \city{Shanghai}
  \country{China}
}
\email{ouyy@shanghaitech.edu.cn}

\author{Shenghan Gao}
\orcid{0009-0008-3397-0341}
\authornotemark[1]
\affiliation{%
  \institution{School of Information Science and Technology \\ ShanghaiTech University}
  \city{Shanghai}
  \country{China}
}
\email{gaoshh1@shanghaitech.edu.cn}

\author{Ruichuan Wang}
\orcid{0009-0007-9897-1954}
\affiliation{%
  \institution{School of Information Science and Technology \\ ShanghaiTech University}
  \city{Shanghai}
  \country{China}
}
\email{wangrch2023@shanghaitech.edu.cn}

\author{Hailiang Zhu}
\orcid{0009-0006-4550-3826}
\affiliation{%
  \institution{School of Information Science and Technology \\ ShanghaiTech University}
  \city{Shanghai}
  \country{China}
}
\email{zhuhl2024@shanghaitech.edu.cn}

\author{Yuheng Shao}
\orcid{0009-0008-6991-6427}
\affiliation{%
  \institution{School of Information Science and Technology \\ ShanghaiTech University}
  \city{Shanghai}
  \country{China}
}
\email{shaoyh2024@shanghaitech.edu.cn}

\author{Xiaoyu Gu}
\orcid{0009-0005-6148-754X}
\affiliation{%
  \institution{School of Creativity and Art \\ ShanghaiTech University}
  \city{Shanghai}
  \country{China}
}
\email{guxy2@shanghaitech.edu.cn}

\author{Quan Li}
\authornote{Corresponding Author.}
\orcid{0000-0003-2249-0728}
\affiliation{%
  \institution{School of Information Science and Technology \\ ShanghaiTech University}
  \city{Shanghai}
  \country{China}
}
\email{liquan@shanghaitech.edu.cn}
\renewcommand{\shortauthors}{Yang Ouyang, Shenghan Gao et al.}

\begin{abstract}
Online comments significantly influence users' judgments, yet their presentation, often determined by platform algorithms, can introduce biases, such as anchoring effects, which distort reasoning. While existing research emphasizes mitigating individual cognitive biases, the evolution of user judgments during comment engagement remains overlooked. This study investigates how presentation cues impact reasoning and explores interface design strategies to mitigate bias. Through a preliminary experiment (N=18) and a co-design workshop, we identified key challenges users face across a four-stage process and distilled four design requirements: pre-engagement framing, interactive organization, reflective prompts, and synthesis support. Based on these insights, we developed \textit{CommSense}, an on-the-fly plugin that enhances user engagement with online comments by providing visual overviews and lightweight prompts to  guide reasoning. A between-subject evaluation (N=24) demonstrates that \textit{CommSense} improves bias awareness and reflective thinking, helping users produce more comprehensive, evidence-based rationales while maintaining high usability.

\end{abstract}



% \ccsdesc[500]{Human-centered computing~Human computer interaction (HCI)}
%%
% \ccsdesc[500]{Human-centered computing~ User interface toolkits}

  % \includegraphics[width=\textwidth]{sampleteaser}
  % \caption{Seattle Mariners at Spring Training, 2010.}
  % \Description{Enjoying the baseball game from the third-base
  % seats. Ichiro Suzuki preparing to bat.}
  % \label{fig:teaser}

\maketitle
\section{Introduction}
\par 
% Today's social media platforms have redefined the role of users, transforming them into active gatekeepers who curate, interpret, and disseminate information~\cite{funk2017americans,li2022identifying}.
% This shift empowers users to actively shape information landscapes, which becomes particularly pronounced when individuals face uncertain, high stakes situations with limited information. In such contexts, they increasingly rely on social platforms to support their decision-making~\cite{mcmullan2019relationships,akhther2022seeking}.
\ouyangre{Online platforms, such as forums and knowledge-sharing sites, have become essential spaces for individuals to gather information, form impressions, evaluate options, and make daily decisions~\cite{li2022identifying,gawer2021online,akhther2022seeking}.
For example, travelers frequently rely on online travel discussions to explore destinations, plan trips, and identify points of interest~\cite{xiang2015information}.
}
% This shift is particularly pronounced when individuals encounter unfamiliar symptoms that evoke health-related anxiety or uncertainty, prompting them to increasingly rely on these platforms for health information~\cite{mcmullan2019relationships,akhther2022seeking}. 
% Unlike traditional sources that are reviewed by medical professionals~\cite{trethewey2019medical,gurler2022assessment}, content on social media—such as user comments—is largely unfiltered and algorithmically prioritized based on metrics such as popularity or recency~\cite{chan2025examiningalgorithmiccurationsocial,Ciampaglia_2018}. 
Unlike traditional information sources (e.g., curated publications or news outlets) where content is vetted by domain experts~\cite{trethewey2019medical,gurler2022assessment}, online platforms lack such safeguards. \ouyangre{Here, user-generated content, including comments, is largely unfiltered and algorithmically prioritized based on metrics such as popularity or recency~\cite{chan2025examiningalgorithmiccurationsocial,Ciampaglia_2018}.} By elevating emotionally charged or sensational content~\cite{brady2020mad}, these algorithmic mechanisms disproportionately expose users to unvetted narratives. Consequently, perceptions become shaped less by clinical accuracy and more by the presentation and sequencing of posts—a dynamic that readily triggers cognitive biases such as the \textit{anchoring effect}~\cite{furnham2011literature,soprano2024cognitive}. 
% For instance, a highly upvoted anecdote describing a rare condition may anchor users' interpretations of their symptoms, causing them to focus on improbable diagnoses while neglecting more likely explanations. Such cognitive distortions can escalate mild curiosity into substantial emotional distress—an outcome particularly concerning in the context of cyberchondria~\cite{peng2021status,schenkel2021conceptualizations,turk2025generalized}.
For instance, an algorithmically amplified comment describing an extreme outcome may anchor users' interpretations—demonstrating how their gatekeeping role (sharing/upvoting) and algorithmic curation jointly distort risk assessment. This leads to overestimation of unlikely scenarios while neglecting probable ones, escalating initial uncertainty into disproportionate concern or indecision, especially when timely decisions are required.
% Moreover, users often conflate visibility with credibility~\cite{pennycook2020falls}, placing undue trust in top-ranked content and thereby increasing their reliance on cognitive shortcuts during judgment.

% Presentation bias in information streams may distort users’ risk assessment by amplifying perceived health risks, escalating curiosity into anxiety, and reinforcing unwarranted confidence in a particular course of action.

% Although prior work in HCI and health informatics has explored the impact of social media on health anxiety and misinformation~\cite{}, most studies focus narrowly on exposure—whether users encounter potentially misleading content. We argue this overlooks a critical dimension: how content is presented.
\par Building on these concerns, while prior research has established that exposure to user-generated content on online platforms can activate a range of cognitive biases, including \textit{availability bias}~\cite{folkes1988availability} and \textit{confirmation bias}~\cite{nickerson1998confirmation}, several critical dimensions remain underexplored. Specifically, existing studies predominantly emphasizes whether users encounter certain types of information~\cite{li2023availability,mamede2010effect,mamede2020immunising}, rather than investigating how the \textit{format} and \textit{sequence} of that information shape users' cognitive judgments. For instance, Li et al.~\cite{li2023availability} modeled public opinions on COVID-19 vaccines and demonstrated that confirmation bias and the availability heuristic—especially when amplified by extreme authoritative sources—can intensify opinion polarization. However, their study does not explore how factors such as presentation order, interface layout, or message salience may moderate these psychological effects. This gap in understanding cognitive modulation also limits intervention strategies. Current approaches primarily target comment content through direct intervention. While manual moderation allows for nuanced assessments~\cite{gongane2022detection}, it is inherently constrained by scalability issues, time delays, and inconsistencies in judgment. Rule-based filtering offers improved efficiency but is often easily circumvented~\cite{gomes2024problematizing} and may inadvertently suppress the authenticity and diversity of user discourse—both of which are essential for reflecting the complex landscape of public health perceptions. 

% \par In response to these challenges, a growing body of work has begun exploring interface-level interventions~\cite{gierth2020attacking,kington2021identifying,ghahramani2022potential}. For example, rendering comments based on source credibility has shown initial promise~\cite{kington2021identifying}. While such interventions offer new pathways, many operate under a static assumption—that users passively absorb credibility cues. In reality, engagement is highly active and mediated by individual differences. Consequently, interventions relying solely on presentation cues often fail to account for how traits like risk perception~\cite{dyer2020public}, emotional state~\cite{yu2020emotions}, and cognitive openness~\cite{ali2025cognitive} act as cognitive filters, modulating how interface cues are interpreted: for instance, users with high-risk sensitivity may disproportionately weight alarming comments regardless of credibility labels, while emotionally aroused individuals might fixate on salient visual cues. Such filtering dynamically modulates whether interface designs mitigate biases or inadvertently reinforce them. Quantifying these personalized interactions is therefore critical: without mapping how traits modulate cue processing, we cannot design robust interventions that adapt to real-world cognitive diversity—particularly in high-stakes, uncertain contexts where biased judgments carry severe consequences.

\par In response to these challenges, a growing body of research has investigated interface-level interventions~\cite{gierth2020attacking,kington2021identifying,ghahramani2022potential}. For instance, displaying comments based on source credibility has proven to be effective~\cite{kington2021identifying}. These strategies typically leverage credibility cues within the interface to assist users in navigating large volumes of content, helping them identify more trustworthy or relevant information~\cite{yang2016meta}. However, such approaches often neglect the dynamic and multifaceted nature of user traits. Individual characteristics—such as risk perception~\cite{dyer2020public}, emotional state~\cite{yu2020emotions}, and cognitive openness~\cite{ali2025cognitive}—are not fixed; they can fluctuate significantly over time. These traits interact with interface cues in complex ways, potentially influencing whether an intervention ultimately mitigates or exacerbates cognitive biases. Therefore, gaining a deeper understanding of these complex interactions is essential for elucidating the mechanisms that shape user judgment and behavior within online platforms.

% Furthermore, few studies have systematically examined how users' self-perceptions—such as risk sensitivity~\cite{dyer2020public}, emotional state~\cite{yu2020emotions}, and tendency to adopt information~\cite{ali2025cognitive}—interact with the presentation style of health information. These individual factors may critically influence how content is interpreted and whether cognitive biases are triggered. Gaining a deeper understanding of these interactions is essential for uncovering the presentational mechanisms that govern user judgment and behavior in online health information environments.

% As an essential prerequisite to quantifying these trait-mediated interactions, we must first systematically establish how core interface cues themselves operate within the complex comment contexts where cognitive diversity matters most.

% As an essential prerequisite, we first need to systematically examine how core interface cues operate within complex comment contexts, particularly where cognitive diversity is high.
\par To explore comment-based sense-making in a controlled yet realistic context, we employ a hotel-booking scenario as a proxy environment. This scenario presents comments with varying sentiment, detail, and reliability, requiring users to interpret, compare, and integrate information. It creates a complex setting, marked by ambiguity, trade-offs, information overload, and the need to balance potentially conflicting cues~\cite{seutter2023sorry}. This approach enables us to systematically investigate how the presentation of comments influences reasoning and bias, while maintaining a familiar and manageable context.

% \par As an essential prerequisite, we first need to systematically examine how core interface cues operate within complex comment contexts, particularly where cognitive diversity is high. To this end, we pose the following foundational research question: 
\par Building on this grounding, we first ask: \textbf{RQ1}: \textit{\ouyang{How does the presentation of comments shape users’ judgment and reasoning as they interpret and integrate complex information on online platforms?}} To examine this, we conducted a between-subjects experiment (Study I, N = 18) in which participants imagined preparing to attend an academic conference in an unfamiliar city and needed to evaluate a hotel. Participants viewed user comments sorted by sentiment valence under one of three conditions: \textit{Positive-First}, \textit{Negative-First}, or \textit{Interleaved}. Using a think-aloud protocol, we traced not only participants' final assessments but also the evolution of their reasoning. Results revealed that users consistently followed a four-stage decision path consisting of \textit{Initial Framing}, \textit{Evidence Foraging}, \textit{Belief Updating}, and \textit{Synthesis and Judgment}, which gave rise to three issues leading to biased judgments: goal alienation from Initial Framing to Verification; asymmetric Evidence–Belief loops driving cognitive oversimplification or overload; and insufficient support for Synthesis Judgment.

\par Building on these findings, we adopted a design-oriented perspective. Understanding these challenges from the user's standpoint was essential for translating their needs into actionable design requirements. This led to our second research question: \textbf{RQ2}: \textit{From the user's perspective, what are the primary usability challenges and key needs when navigating complex comment environments, and what design strategies can mitigate biased perceptions?} To answer this, we conducted a follow-up co-design workshop (Study II) with the same 18 participants from Study I. Participants reflected on moments of confusion and perceived bias during their decision-making process, and collaboratively proposed design concepts for potential solutions. Through this user-centered process, we distilled their insights into four core design requirements: (1) pre-engagement framing, (2) interactive organization and annotation, (3) in-situ reflective prompts, and (4) dynamic synthesis support.

\par Guided by these requirements, we designed \textit{CommSense}, a lightweight tool that supports more rational and reflective engagement with online comments. The interface integrates three coordinated components: the \textit{Topic Corpus Overview}, \textit{Comment Navigation Panel}, and \textit{Synthesis Board}, to scaffold a continuous sensemaking process. Together, these components help users grasp the overall discourse, detect trends, explore contrasting perspectives, and annotate evidence. By consolidating reflections and curated insights, \textit{CommSense} aims to facilitate synthesis, mitigates bias, and promotes balanced evaluation. This leads to our third research question: \textbf{RQ3}:  \textit{To what extent can interface-level strategies reduce bias induced by comment presentation?} To answer this, we conducted a between-subjects experiment (Study III, N=24) comparing \textit{CommSense} with a simplified, conventional baseline interface. Results showed that \textit{CommSense} enhanced users' bias awareness and reflective thinking, enabling them to generate more comprehensive, evidence-based rationales while maintaining high usability. Notably, all studies reported above (I, II, and III) were approved by the Research Ethics Committee and conducted in accordance with standard ethical guidelines. In summary, the primary contributions of this work are as follows:

\begin{itemize}[itemsep=0.5ex, topsep=1ex]
    \item \ouyang{We first investigated how users form judgments in complex comment environments (\textbf{RQ1}, Study I), finding a four-stage decision path—\textit{Initial Framing}, \textit{Evidence Foraging}, \textit{Belief Updating}, and \textit{Synthesis and Judgment}—with goal alienation, asymmetric evidence–belief loops, and insufficient synthesis leading to biased outcomes.} 
    \item \ouyang{We designed \textit{CommSense} (\textbf{RQ2}, Study II), a lightweight tool that scaffolds structured engagement through pre-engagement framing, interactive organization and annotation, reflective prompts, and dynamic synthesis support to mitigate bias.}
    \item \ouyang{We conducted a user study (\textbf{RQ3}, Study III) that demonstrated how \textit{CommSense} enhances bias awareness, supports reflective reasoning, and fosters evidence-based judgments, all while maintaining high usability.}
\end{itemize}

\section{Related Work}
% \subsection{Information-Seeking Support in Online Social Media}
\subsection{Information-Seeking and Sensemaking Support on Online Platforms}

% \ouyangre{
% Online platforms, including forums and knowledge-sharing sites, host substantial volumes of user-generated content that people routinely draw on for information seeking and decision- making~\cite{wu2016survey,mishra2015information}.
\ouyangre{Online platforms, including forums and knowledge-sharing sites, host large volumes of user-generated content that people often rely on for information seeking and decision-making~\cite{wu2016survey,mishra2015information}. Many studies show that users’ evaluation and interpretation of content are shaped by cognitive and affective factors~\cite{metzger2003college,sundar2008main}. For example, Wilson’s model of information behaviour~\cite{wilson1999models} explains how these factors are influenced by personal, social, and environmental contexts, which in turn interact with individual differences in identifying, weighing, and integrating information~\cite{mishra2015information}. Emotional responses further modulate judgments, highlighting the interplay of cognition, context, and affect~\cite{arapakis2008affective}.}

\ouyangre{In practice, the content’s scale, unstructured nature, and variable quality make it challenging for users to identify information that is both relevant and trustworthy~\cite{chowdhary2023systematic}. To support users in making sense of large-scale content and forming informed judgments, a range of approaches has been developed. Recommender systems, as a relatively early and foundational approach, are a primary method for managing overload by suggesting content based on preference modeling and choice architecture~\cite{jameson2022individual, schnabel2016using}. However, these techniques often fall short in helping users understand a corpus’s underlying structure or the diversity of perspectives it contains~\cite{pariser2011filter,mcnee2006being}.}
\par \ouyangre{Alongside algorithmic work, recent advances in visualization and interaction aim to uncover the underlying structure and patterns in information, facilitating more effective sense‑making~\cite{cao2015targetvue,viegas2004newsgroup,reda2011visualizing}. For instance, VisOHC~\cite{kwon2015visohc} visualizes discussion threads as collapsible boxes, allowing users to quickly assess information structure and quality. ConVis~\cite{hoque2014convis} clusters topics and opinions from complex discussions into an interactive visual interface to streamline exploration. Similarly, OpinionFlow~\cite{wu2014opinionflow} combines Sankey diagrams with density maps to visualize the propagation and evolution of opinions, while the more recent ComViewer~\cite{wu2025comviewer} organizes discussions and highlights critical exchanges to help users identify key information.}

\par \ouyangre{Existing tools highlight the potential of innovative approaches to enhance information seeking on online platforms. By moving beyond conventional ranked lists, these systems enable users to better navigate and make sense of complex, large-scale datasets, effectively addressing issues of information overload and disorganization. However, they primarily focus on organizing and visualizing comment content, often overlooking how its presentation dynamically shapes the user's underlying cognitive journey. Our work addresses this gap (\textbf{RQ1}) by systematically investigating how different presentation strategies influence this judgment formation process in Study I, tracing users’ sensemaking trajectories to reveal bias emergence and providing insights to guide the design of interface-level interventions that foster more reflective and rational judgments.}

\subsection{Bias Detection and Mitigation on Online Platforms}
\label{sec:bias_re}
% Contradict with your writing on notion

% \par \ouyang{Health-related content on social media is often presented without medical oversight and varies widely in tone, completeness, and credibility, which can make it difficult for users to form accurate perceptions. This environment reduces the opportunity to develop stable intuitions or discern meaningful patterns, leaving users vulnerable to cognitive biases such as the anchoring effect~\cite{}, attraction effect~\cite{}, confirmation bias~\cite{} and availability bias~\cite{}. These biases have been extensively studied and categorized in comprehensive taxonomies~\cite{}.} 
\par User-generated content on online platforms often lacks expert oversight and exhibits considerably variability in tone, completeness, and credibility. This inconsistency makes it difficult for users to form accurate judgments or identify reliable patterns, thereby increasing susceptibility to cognitive biases. In such unstructured environments, users are particularly prone to well-documented biases including anchoring~\cite{cho2017anchoring,pohl2004cognitive,selim2021anchoring,valdez2017priming}, the attraction effect~\cite{dimara2018mitigating}, confirmation bias~\cite{burke2005improving}, and availability bias~\cite{tversky1973availability}. These cognitive pitfalls have been systematically studied and organized in taxonomies that map their influence on human decision-making~\cite{dimara2018task,nussbaumer2016framework,pohl2014sensemaking,wall2017warning}. Recent research further shows that even subtle changes in the order of information presentation can significantly shape decisions~\cite{akl2016designing}, and that prior judgments may act as cognitive anchors, resulting in errors or inconsistencies when reassessing the same scenario~\cite{echterhoff2022ai}.
% Among these, availability bias and confirmation bias are particularly relevant in our context. 
\par The pervasive influence of cognitive biases on online platforms can be meaningfully interpreted through the lens of dual-process theory. This framework posits that human cognition operates via two distinct systems: \textit{System 1}, which is fast, intuitive, and heuristic-driven, and \textit{System 2}, which is slower, more deliberate, and analytical~\cite{kahneman2011thinking}. When engaging with the rapid, high-volume flow information in online environments, users often rely on \textit{System 1} processing as a cognitively efficient strategy~\cite{bago2020fake}. Although generally adaptive, this reliance increases vulnerability to systematic errors when intuitive judgments override careful analysis~\cite{bago2020fake}. Availability bias and confirmation bias are among the most prominent \textit{System 1}-driven biases. Availability bias arises when individuals overestimate the relevance or frequency of an event simply because it is easily recalled~\cite{evans2013dual}. In the context of online platforms, repeated exposure to similar content enhances the familiarity and salience of certain topics, causing users to overweight them in their judgments~\cite{luzsa2021false}. This effect is further amplified by algorithmic curation, which prioritizes content based on prior engagement patterns, reinforcing perceived importance and contributing to the formation of implicit social norms~\cite{burton2024simple, shin2022countering}. Confirmation bias, on the other hand, reflects the tendency to seek, interpret, and remember information in ways that confirm one's existing beliefs~\cite{nickerson1998confirmation}. Online platforms exacerbate this bias through algorithmic personalization and selective exposure, resulting in highly curated content streams that validate users' prior views. Over time, this leads to the formation of echo chambers where disconfirming information is filtered out or dismissed~\cite{luzsa2021false, putri2024echo}, making users more resistant to alternative perspectives and more susceptible to entrenched misconceptions~\cite{echterhoff2022ai}.

\par Detecting and mitigating cognitive bias remains a critical concern in supporting sound decision-making, particularly in high-stakes or time-constrained scenarios. Prior research in this area generally falls into four broad categories: 1) \textit{Prevention-based approaches} aim to reduce bias either by enhancing users’ awareness or by modifying the decision-making environment to foster more reflective thinking. A widely adopted approach involves targeted training aimed at raising awareness of cognitive biases and fostering deliberate reasoning~\cite{fischhoff1975silly,capers2017implicit}. Although this method can be effective, such training-based interventions often impose a substantial cognitive burden and depend heavily on users' prior knowledge and motivation to actively participate. Alternatively, bias mitigation can be integrated seamlessly into existing workflows without explicitly mentioning bias, for example, by enhancing transparency~\cite{zhu2022bias} or improving the accessibility of relevant information~\cite{talkad2018making}, thereby facilitating more informed and accurate decision-making.
2) \textit{Detection methods} aim to identify the presence of bias during the decision-making process. These approaches frequently employ machine learning models or interactive visual interfaces to surface patterns of biased behavior~\cite{nussbaumer2016framework,sinha2022personalized}. Some studies propose measurable indicators~\cite{wall2019formative,wall2017warning} that make cognitive biases more visible and actionable. 3) \textit{Localization strategies} seek to determine the specific moments and conditions under which bias arises. By visualizing contextual cues and decision traces, these methods provide a more fine-grained understanding of how biases operate in situ~\cite{wall2021left,narechania2021lumos}. For instance, Echterhoff et al.\cite{echterhoff2022ai} developed a probabilistic model that successfully identified anchoring effects in peer-review decisions. 4) \textit{Mitigation strategies} are designed to directly reduce the influence of bias on outcomes. These include strategies such as reordering information to minimize framing effects~\cite{akl2016designing}, leveraging visual tools like design spaces~\cite{wall2019toward}, or incorporating graphical cues to encourage deeper reflection~\cite{chuanromanee2022crowdsourced}. Empirical studies have demonstrated that embedding such visual aids into decision workflows can meaningfully reduce bias and ease users' cognitive burden~\cite{sukumar2018visualization,talkad2018making,liu2024biaseye,boonprakong2025hci}.

% \par Drawing on prior research, we investigate how user-generated comments on online platforms shape biased perceptions and explore strategies to address them (\textbf{RQ2}, Study II). 
\par \ouyangre{Drawing on prior research, we investigate how user-generated comments on online platforms shape biased perceptions and explore strategies to address them, as examined in Study II for \textbf{RQ2}.} Our approach \textit{prevents} initial anchoring through pre-engagement framing, \textit{detects} and \textit{localizes} bias via in-situ reflective cues, and \textit{mitigates} it through synthesis support. These interventions serve as subtle \textit{nudges}~\cite{caraban201923}, encouraging users to engage more deliberately and form reasoned judgments.

\subsection{Techniques for Comment Presentation and Visualization}

\par To help users navigate the large volume of user-generated content, HCI researchers have developed a variety of techniques for content presentation and visualization~\cite{chen2024amplifying,liu2023coargue,ormel2021using}. Some tools focus on filtering and structuring content using sentiment analysis and relevance ranking to improve information accessibility and user experience. For example, \textit{CoArgue}~\cite{liu2023coargue} organizes claims in Q\&A forums based on sentiment and relevance, streamlining users' ability to locate pertinent information. Similarly, Chen et al.~\cite{chen2024amplifying} employed emotion analysis to song-related comments, enhancing users' emotional engagement and fostering empathy. Other systems aim to support content comprehension and user engagement. Some platforms automatically highlight key comments~\cite{hoque2017cqavis,liu2022planhelper}, while interactive note-taking tools enable users to capture and annotate information~\cite{bauer2006evaluating,cao2022videosticker,fang2021notecostruct,kang2021metamap}. Social annotation tools further enable collaborative reflection by allowing users to annotate and discuss shared content~\cite{almahmoud2024enhancing,gao2013case,hwang2007study}. For instance, Hwang et al.~\cite{hwang2007study} introduced features such as text highlighting, annotations, and voice recording, which were found to support more focused and meaningful discussions than traditional online forums~\cite{brush2023supporting}.

\par These systems are underpinned by a range of natural language processing (NLP) techniques such as text summarization~\cite{keikha2014evaluating,song2017summarizing}, classification~\cite{peng2021effects,sharma2018mental,wambsganss2020conversational,yang2019seekers,yang2019channel}, and clustering~\cite{peng2020exploring,krause2017critique}, which are used to extract, categorize, and structure comment data. In design-related communities, for example, Guo et al.~\cite{guo2023makes} analyzed feedback through lenses such as actionability and specificity, while Peng et al.~\cite{peng2024designquizzer} used fine-tuned language models and clustering techniques to organize comments based on sentence type and UI-related keywords. These approaches provide a foundation for fostering reflective and informed interaction with complex content. 
% Building on this foundation, our study explores how restructuring comment presentation can heighten users' awareness of cognitive biases and reduce their impact on perception and judgment.
\ouyangre{Building on this foundation, our study explores how restructuring comment presentation can heighten users’ awareness of cognitive biases and reduce their impact on perception and judgment, which we evaluate in Study III (\textbf{RQ3}).}

% \section{Methodology}
\section{Study I: Investigating the Impact of Comment Presentation Strategies on Biased Perceptions}
\label{sec:studyI}
\par
\begin{figure*}[h]
    \centering
    \includegraphics[width=\linewidth]{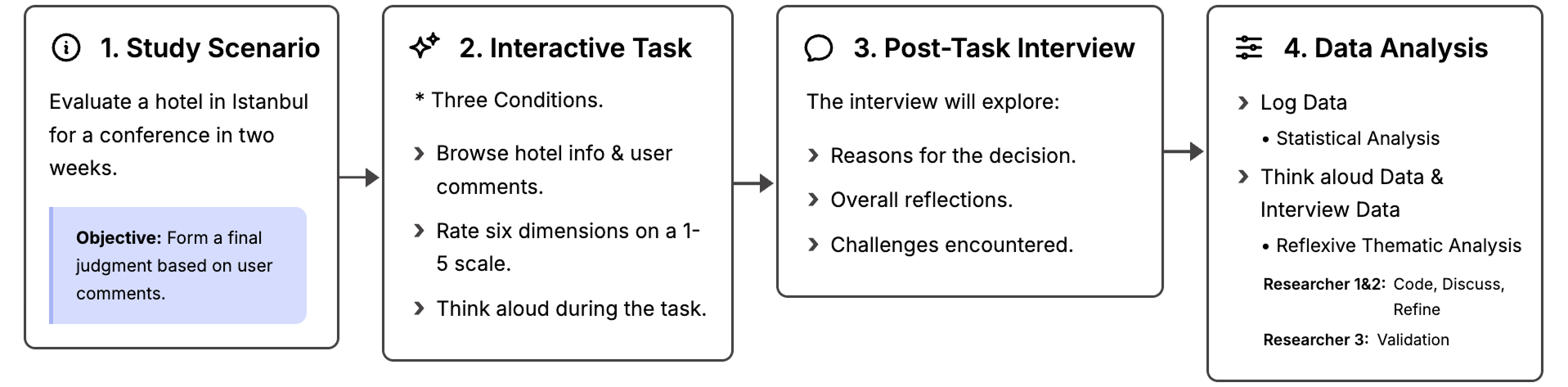}
    % \caption{An overview of Study I, including the scenario, task, procedure, and data analysis method.}
    \caption{The study began by having participants evaluate a hotel in Istanbul based on user comments under three conditions, thinking aloud and rating it on a 1–5 scale. Post-task interviews and interaction logs were analyzed using statistical methods and reflexive thematic analysis, with results validated by a third researcher.}
    \label{fig:studyI}
\end{figure*}

 To address \textbf{RQ1}, we conducted Study I, with an overview of the study design and procedure presented in \cref{fig:studyI}.

\subsection{Participants}
\par We recruited 18 graduate students (10 males, 8 females; aged 20–28) from a local university via online bulletin boards and word of mouth to participate in the study. All participants had prior experience booking hotels online and routinely consulted user-generated comments when making hotel-related decisions. Their information are shown in \cref{sec:ParticipantsinstudyI}. All data were anonymized: only participant pseudonyms were retained, and no personally identifiable information was recorded or stored.

\subsection{Research Sites and Data Preparation}
% \subsection{Data Preparation}
\label{sec:dataset}
\par To place participants in realistic hotel evaluation scenarios, we first collected authentic user reviews from leading travel platforms, e.g., \textit{Booking}, \textit{TripAdvisor}, \textit{Agoda}, and \textit{Triping}, chosen for their extensive, credible content that offer rich insights into guest experiences. From this collection, we curated high-quality comments to serve as experimental stimuli. Drawing on established guest satisfaction dimensions~\cite{li2020comprehending,de2022hotel}, we focused on six key criteria: room (quality and amenities), service (staff professionalism and responsiveness), value (cost-effectiveness), location (proximity and accessibility), sleep quality (comfort), and cleanliness (hygiene). To ensure relevance and clarity, we excluded comments unrelated to these aspects or containing off-topic content such as travel itineraries and photographs, resulting in a reliable dataset that captures both the tangible and experiential facets of hotel stays.
% \par A key challenge in such research is the inherent complexity of online comments. To isolate the specific impact of comment style, our pre-study focused on distinguishing and standardizing two fundamental categories: factual and subjective comments. This distinction is critical, as a body of research has demonstrated that these styles differentially shape perceived trustworthiness~\cite{}, with factual language enhancing objective credibility and subjective language introducing emotional ambiguity~\cite{}. By empirically validating our stimuli, we could ensure robust experimental control in the subsequent main study.

\subsection{Experimental Conditions}
\par After filtering for high-quality comments, we conducted further analysis and refinement of the comment dataset. On real-world platforms, comment order is typically determined by proprietary algorithms that weigh factors such as recency or popularity. These algorithms are often complex and non-transparent, making them difficult to replicate in controlled experimental settings. Rather than attempting to reproduce such algorithms, our study isolates and examines the effect of comment ordering itself—specifically, how the relative position of comments influences user perception and decision-making.
\par Building on established research highlighting the influence of sentiment polarity on user trust and attention~\cite{burke2005improving}, we adopted comment valence (positive or negative) as the primary sorting criterion. To classify sentiment at scale, we employed a few-shot approach using OpenAI's GPT-4o, chosen for its advanced contextual understanding and operational flexibility. Compared to fine-tuned discriminative models like RoBERTa~\cite{liu2019roberta}, this LLM-based method enables effective analysis without task-specific training data~\cite{brown2020language,wei2022emergent}, offering superior adaptability for large and evolving text corpora. In practice, GPT-4o performed initial automated annotations using prompts containing representative positive and negative examples. To ensure accuracy, two researchers independently reviewed and corrected these classifications. Discrepancies were resolved through discussion to establish consensus labels. This hybrid approach, combining automated efficiency with expert human oversight, allowed for the efficient processing of large volumes of content while maintaining labeling consistency, preserving quality, and capturing nuanced expressions such as sarcasm and subtle sentiment. Based on this sentiment classification, we manipulated comment arrangement to examine its impact on user perception by designing three valence-based presentation conditions:
% Based on this sentiment classification, we experimentally manipulated comment arrangement to examine its impact on user perception. Specifically, we designed three presentation conditions, each defined by a different valence-based ordering:

\begin{itemize}[itemsep=0.5ex, topsep=1ex]
    \item \textbf{Positive-First:} All positive comments appear at the top of the list, followed by all negative comments.
    \item \textbf{Negative-First:} All negative comments appear at the top of the list, followed by all positive comments.
    \item \textbf{Interleaved (Control):} Positive and negative comments are presented in a strictly alternating sequence throughout the list.
\end{itemize}

\par For each condition, six participants were randomly selected from the overall participant pool. This design enabled a systematic investigation of how comment order influence user perceptions and allowed us to explore potential primacy and negativity biases in decision-making.

\subsection{Experimental Task and Procedure} 
% We adapted our study design from~\cite{jiang2023graphologue,yuan2023critrainer,chen2025redesign} to structure the task and procedure. 
\par We adapted our main study design from prior studies~\cite{jiang2023graphologue,yuan2023critrainer,chen2025redesign}, tailoring the task and procedure to simulate a realistic hotel assessment scenario in which users read comments to form judgments. The study began with an introduction to the task contexts:
\begin{quote}
``Imagine you are planning to attend an academic conference in an unfamiliar city (e.g., Istanbul) in two weeks. While browsing a hotel booking platform, you find a hotel that seems like a good fit based on its description. To evaluate it more thoroughly, you decide to read through the user comments.''
\end{quote}
\par Rather than asking participants to make a definitive booking decision, our study focused on how individuals evaluate a single hotel based on user-generated comments—a process that mirrors common real-world behavior, where people aim to form a well-rounded impression of one specific option rather than choose among multiple alternatives. Comparative decision-making often entails more complex cognitive demands, such as weighing trade-offs across options and adjusting decision strategies~\cite{busemeyer2019cognitive,gluth2020value}, which fall outside the scope of our study.

\begin{figure*}[h]
    \centering
    \includegraphics[width=\linewidth]{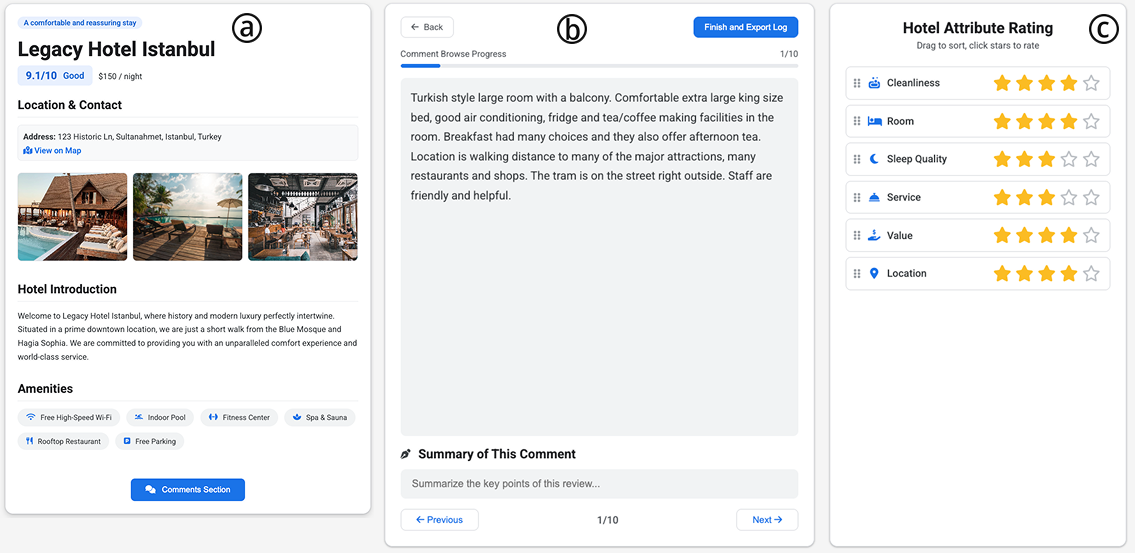}
    \caption{Experimental Probe. Participants first viewed the (a) Hotel Overview Panel, then browsed user reviews in the (b) Review Browsing Interface, and rated six hotel attributes in the Attribute Rating Panel (c) based on the review content. The interface supported iterative evaluation, allowing participants to revise their ratings as their impressions evolved during the review browsing process.}
    \label{fig:ex_interface}
    % \vspace{-4mm}
\end{figure*}

\par Following the scenario introduction, participants interacted with a simulated experimental probe (\cref{fig:ex_interface}) composed of three main components: the Hotel Overview Panel (\cref{fig:ex_interface} (a)), the Review Browsing Interface (\cref{fig:ex_interface} (b), and the Attribute Rating Panel (\cref{fig:ex_interface} (c)). The Hotel Overview Panel presented essential information about the hotel, including its name, price range, representative images, a brief description, and listed amenities, offering participants a concise overview. Participants were randomly assigned to one of three conditions without being informed of the sorting criteria. The assignment was balanced to ensure equal participant distribution across conditions. Subsequently, participants browsed a curated selection of user-generated reviews within the Review Browsing Interface, adapted from our dataset (\cref{sec:dataset}), to gather evaluative information.

\par Simultaneously, the Attribute Rating Panel allowed participants to assess the hotel across six key dimensions—room quality, service, value, location, sleep quality, and cleanliness—using 5-point Likert scale sliders (1 = very poor, 5 = excellent). Participants were free to switch between browsing reviews and adjusting their ratings, facilitating an iterative evaluation process that reflected their evolving impressions.

\par Participants were given 15 minutes to engage with the probe. Upon completing the browsing task, they provided a brief overall evaluation in 1–3 sentences, referencing the information they had reviewed. 
% This was followed by semi-structured interviews aimed at eliciting reflections on their experience, with particular focus on shifts in attention and changes in perception throughout the information-gathering process. 
% Interview prompts included questions such as, ``\textit{How did your attention shift while browsing posts and comments?}'' and ``\textit{Did you perception of the hotel change or become uncertain as you reviewed the content?}''
\ouyangre{This was followed by semi-structured interviews aimed at eliciting reflections on participants' experience. The semi-structured interviews were designed by three authors, and refined through pilot testing to ensure clarity and alignment with study objectives. The questions were open-ended and flexible, allowing researchers to follow participants’ responses and probe specific behaviors. For instance, if a participant mentioned a shift in attention, the interviewer would ask \textit{``Can you describe how your attention shifted?''} or \textit{``What prompted this change in focus?''}. The interview protocol was shaped to explore pre-defined question points, including attention allocation, evaluation strategies, moments of reconsideration, and changes in perception, while maintaining openness to participants’ perspectives. Additional follow-up prompts encouraged participants to elaborate on specific actions, for example: \textit{``Which posts or comments stood out to you, and why?''} and \textit{``How, if at all, did your perception of the hotel change as you reviewed the content?''}. This approach ensured that the questions functioned as flexible prompts to guide participants in describing their own experiences without suggesting or directing specific answers.
}
Each interview lasted approximately 15 minutes, with the entire session taking about 40 minutes. Participants received monetary compensation of approximately \$6 USD for their time. 
% \begin{wrapfigure}

\subsection{Data Collection and Analysis}
\begin{figure*}[h]
    \centering
    \includegraphics[width=\linewidth]{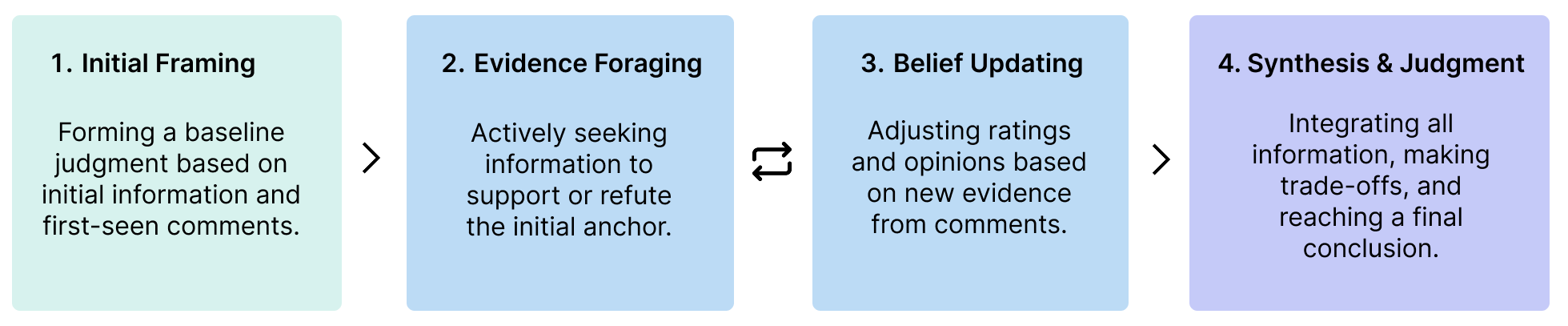}
    \caption{The decision path progresses through four stages: 1) Initial Framing, followed by an iterative cycle of 2) Evidence Foraging and 3) Belief Updating, and concluding with 4) Synthesis and Judgment.}
    % \vspace{-0.5cm}
    \label{fig:decision_path}
\end{figure*}
\par The entire study session was video-recorded and transcribed using \textit{Otter.ai}\footnote{\url{http://Otter.ai}} for subsequent analysis. The collected data comprised two components: (1) interaction log data from the experimental probe, and (2) verbal data from think-aloud protocols and semi-structured interviews. Our analysis of the interaction logs focused on two key aspects of user behavior: 1) the \textbf{Rating Change History} captured every modification participants made to the star ratings of individual hotel attributes (e.g., room, service, location). Each log entry recorded the specific attribute, the updated score, a timestamp, and the review being viewed at the time, enabling us to trace how participants' evaluations evolved in response to different information. 2) the \textbf{Final Attribute Ratings and Ranking} documented participants' final star ratings for each attribute alongside the ultimate order in which they ranked these attributes through drag-and-drop. Together, these data provide a comprehensive view of participants' concluding judgments, reflecting both their assessment of the hotel's qualities and their personal prioritization of attribute importance.
% \par The verbal data were transcribed and analyzed using reflexive thematic analysis~\cite{braun2012thematic}. Two authors independently coded the transcripts and then collaboratively refined the codes to ensure consistency and comprehensiveness. A third researcher subsequently reviewed and validated the final themes. 

\par All verbal data from the interviews were transcribed and prepared for analysis. A reflexive thematic analysis~\cite{braun2012thematic} was conducted in four iterative phases: code generation, theme construction, revising, and refining. In the initial phase, the first and second authors selected two transcripts from each participant group (12 total) to develop a provisional set of inductive codes. These codes were treated as data-driven, semantic labels reflecting participants' expressed meanings, not as predefined categories, with the expectation they would evolve through reflexive analysis. Using NVivo 14~\cite{allsop2022qualitative}, they independently coded the transcripts, documenting their reflections, coding strategies, and interpretations. Through discussion and reflexive consideration, the codes were refined, redundancies were reduced, and related concepts were grouped, resulting in a set of 66 codes.
\par Subsequent coding of the remaining transcripts was conducted iteratively, with the authors regularly meeting to discuss emerging codes, resolve ambiguities, and refine the coding scheme. A third researcher later reviewed the coding process and contributed reflexive insights to the interpretation of the final themes. Throughout this process, coding decisions were guided by reflexive engagement with the data rather than inter-rater reliability~\cite{braun2006using}, and themes were iteratively revised against the full dataset to ensure coherence and depth. This iterative development of themes was not a product of staged coding procedures (e.g., axial or selective coding). Instead, it emerged through the researchers' ongoing process of interpreting, comparing, and reconfiguring codes in light of the entire dataset. Finally, the process yielded a final set of 53 codes, as listed in \cref{sec:codebook}.
\par Interaction log data were analyzed using descriptive statistical methods to characterize participants’ engagement patterns and task-related behaviors. To enable a richer interpretation of user intentions, the log and verbal data were synchronized through timestamp alignment.

% Interviews were audio-recorded, transcribed verbatim, and analyzed using reflexive thematic analysis~\cite{braun2012thematic}.

\subsection{Results}
\label{sec:reofstudyI}
\par We first introduce a universal four-stage decision path observed among all participants. We then examine how participants' cognitive trajectories diverged across experimental conditions, leading to the identification of three decision-making trajectory issues that underpin our key findings.

\subsubsection{A Four Stage Decision Path}
\par To understand the complexity of users' decision-making, we first identified a decision path common to all experimental conditions. Analysis of behavioral logs and think-aloud protocols from 18 participants revealed a consistent four-stage process: \textit{Initial Framing}, \textit{Evidence Foraging}, \textit{Belief Updating}, and \textit{Synthesis \& Judgment} (see \cref{fig:decision_path}). The process begins with \textit{Initial Framing}, in which participants form a preliminary judgment based on the earliest information encountered. This initial impression often serves as a baseline for subsequent information gathering and evaluation, with participants frequently drawing on salient cues such as prominent comments, key ratings, or highlighted features to quickly establish an initial perspective~\cite{luzsa2021false}.

\begin{figure*}[t]
    \begin{center}
    \includegraphics[width=\linewidth]{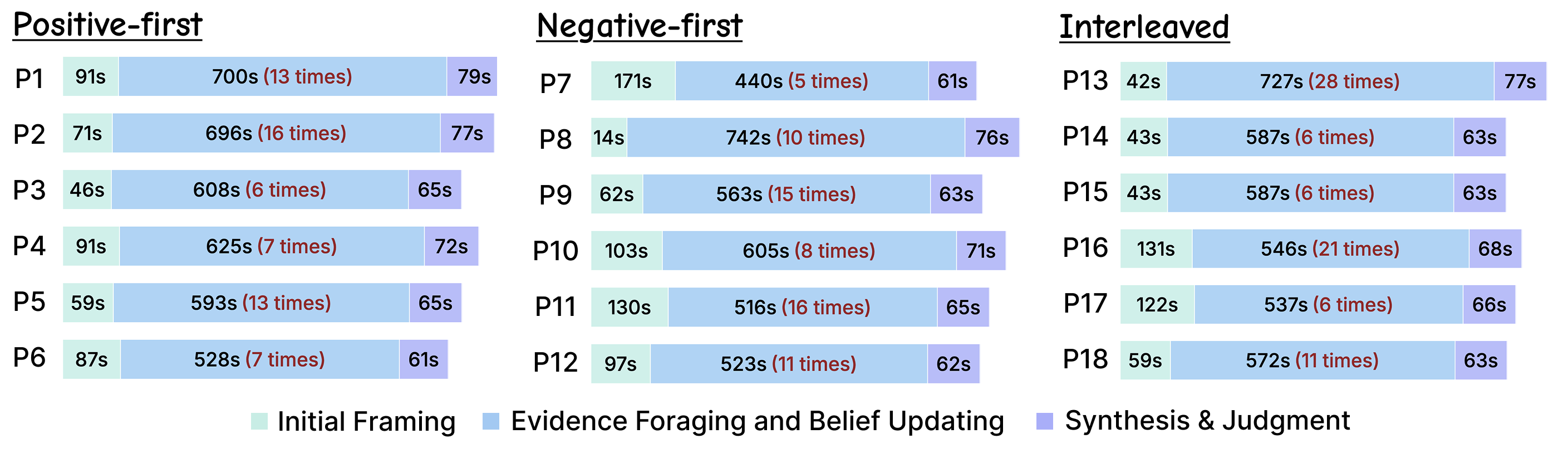}
    % \vspace{}
    \caption{\ouyang{The four-stage decision paths of all 18 participants over time, grouped by experimental condition. Each horizontal bar represents one participant’s session, with colored segments indicating the time spent in each stage.  The numbers in parentheses denote the frequency of score adjustments during Stages 2 and 3.}}
    \label{fig:databaseUserTable}
  \end{center}
\end{figure*}

\par Next, participants progress to the \textit{Evidence Foraging} stage, characterized by an active and often strategic search for supplementary information. During this phase, they scan user comments to identify evidence that either supports or challenges their initial framing. Foraging behaviors vary: some participants selectively focus on comments aligning with their preliminary views, while others deliberately seek out contradictory perspectives. The depth and breadth of their searches reflect individual differences in uncertainty tolerance and information processing styles. Closely intertwined with this is the \textit{Belief Updating} stage, in which participants iteratively revise their opinions and ratings as new evidence emerges. Rather than unfolding in a simple linear sequence, \textit{Evidence Foraging} and \textit{Belief Updating} form a dynamic, cyclical loop: Participants repeatedly alternated between seeking additional comments and updating their beliefs, frequently reassessing earlier conclusions in light of new insights—sometimes reinforcing, sometimes revising their views. This iterative interplay highlights the adaptive and evolving nature of the decision-making process. 
\par Finally, in the \textit{Synthesis \& Judgment} stage, participants consolidate the information gathered to reach a final conclusion. This stage involves weighing conflicting evidence, resolving uncertainties, and forming a judgment that reflects their overall evaluation. 
\par The temporal dynamics of this four-stage decision path, along with the relative time spent in each stage, are visualized in \cref{fig:databaseUserTable}. Each horizontal bar represents the full session of one of the 18 participants. Stage boundaries were defined by key user actions: the \textit{Initial Framing} stage concludes once a participant submits their first complete set of ratings; the combined \textit{Evidence Foraging} and \textit{Belief Updating} stage extends from that point until the final comment is processed; and the subsequent period—dedicated to reviewing notes and finalizing conclusions—constitutes the \textit{Synthesis \& Judgment} stage. In this way, \cref{fig:databaseUserTable} offers a macroscopic, data-driven view that substantiates the common decision path we describe.
\par While participants differed in rating trajectories and underlying rationales, the consistent presence of these four stages across conditions points to a shared cognitive pathway—one that forms the analytical foundation for our subsequent analysis.

% This is followed by Evidence Foraging, an active search for additional information to confirm or challenge the initial impression. Foraging is closely intertwined with Belief Updating, during which participants iteratively adjust their judgments in response to new evidence. Finally, in the Synthesis \& Judgment stage, users integrate the accumulated information to reach a conclusive decision. While participants varied in their specific rating trajectories and booking rationales, the presence of these four core stages was consistent, indicating a shared cognitive pathway across conditions. This four-stage model provides the analytical framework for our subsequent analysis.

\subsubsection{Cognitive Unfolding Along the Four-Stage Path.}
\par In this section, we elaborate on how participants' cognitive processes unfold across the four stages—\textit{Initial Framing}, \textit{Evidence Foraging}, \textit{Belief Updating}, and \textit{Synthesis \& Judgment}.

\par \textit{\textbf{Early Anchoring through Initial Framing.}} Nearly all participants (17/18) completed their preliminary attribute ratings and rankings after reading just the first two comments. In the \textit{Positive-First} condition, early exposure to favorable cues often led to strong initial impressions and high confidence, leaving little room for doubt or reconsideration. As P3 and P4 described, the early positive cues felt ``\textit{pleasant enough}'', with P3 vividly adding that he ``\textit{want[ed] to lounge by the pool and enjoy the beauty.}'' In contrast, the \textit{Negative-First} condition established an immediate unfavorable impression, lowering expectations and triggering a sense of unease. This discomfort prompted participants to adopt a more cautious, scrutinizing approach when processing subsequent information. As P7 reflected, I ``\textit{started off feeling uncertain and had to be more careful about what to believe,}'' signaling a more guarded and deliberative stance. The \textit{Interleaved} condition, however, disrupted the formation of any stable anchor. Ratings for different attributes clustered around ``3'', with minimal variation. The immediate juxtaposition of positive and negative perspectives heightened uncertainty from the outset, fostering sustained open-mindedness and encouraging a more effortful yet comprehensive evaluative process.

\begin{figure*}[h]
    \centering
    \includegraphics[width=\linewidth]{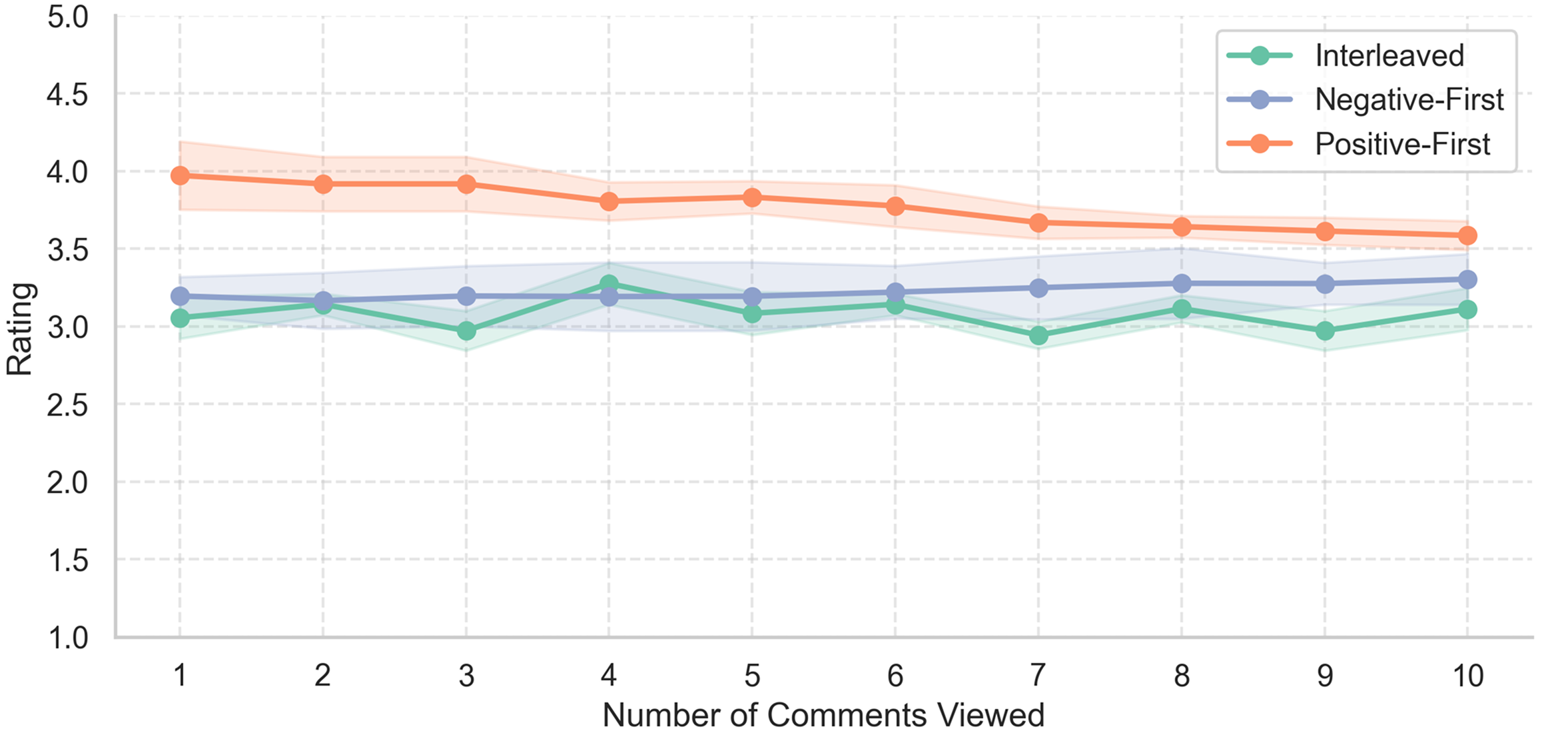}
    \caption{\ouyangre{Average rating trajectories across the three conditions. Lines represent the mean rating given by participants after viewing each comment, with shaded areas indicating $\pm 1$ standard deviation across participants. Positive-First (orange) starts higher, Negative-First (blue) starts lower, and Interleaved (green) shows moderate, less variable ratings. The x-axis indicates the number of comments viewed (1–10), and the y-axis shows the 1--5 rating scale.}}
    \label{fig:rating_trajectories}
\end{figure*}

% This early confidence translated into only light engagement with subsequent stages, as many felt no pressing need to question or expand upon their first impressions.
% \par \textit{\textbf{Early Anchoring through Initial Framing.}}
% Participants in the \textit{Positive-First} condition tended to form favorable impressions early, with limited foraging or revision thereafter. Several participants (e.g., P2, P5) indicated that early positive cues were "reassuring enough," leading to shorter decision paths. In contrast, those in the \textit{Negative-First} condition engaged in prolonged foraging and iterative belief revision. For example, P10 reported actively "looking for balance" to offset the initially negative framing. The \textit{Interleaved} group demonstrated the most distributed engagement, often toggling between conflicting cues and making multiple rounds of micro-adjustments before reaching a conclusion.
% On average, participants adjusted their ratings xx times per session, but the pattern of these changes varied significantly across conditions.

\par \textit{\textbf{Dynamics of Evidence Foraging and Belief Updating.}} To examine how participants' evidence-gathering behaviors and beliefs evolved beyond their initial impressions, we tracked changes in hotel attribute ratings throughout the task. 
\ouyangre{
As shown in \cref{fig:rating_trajectories}, the figure presents the mean rating provided by participants after viewing each comment across the three conditions, with shaded areas indicating ±1 standard deviation. Ratings in the Positive-First condition (orange) start higher, the Negative-First condition (blue) start lower, and the Interleaved condition (green) remain moderate with less variation, illustrating alignment between quantitative trends and participants’ subjective responses.}

Participants in the \textit{Negative-First} condition often delayed rating adjustments until later stages, indicating a cautious, deliberate approach \ouyangre{(see \cref{fig:rating_trajectories}, blue curve)}. For instance, P8 raised their ``Service'' score from ``2'' to ``4'' after reading the final comment, noting ``\textit{At first, I was influenced by the early comments, but the later ones changed my impression. I believe the description in the early comments was likely just a rare occurrence.}'' On the other hand, \textit{Positive-First} participants tended to finalize their ratings early, making only minor downward adjustments \ouyangre{(see \cref{fig:rating_trajectories}, orange curve)}. As P6 remarked, ``\textit{I liked what I saw early on, and that didn't change much.}'' Similarly, P3 commented, ``\textit{I can understand why others might feel uncomfortable, but I believe I wouldn't be affected even if I were in their situation.}'' Participants in the \textit{Interleaved} condition exhibited the most dynamic rating trajectories, frequently revising scores in both directions \ouyangre{(see \cref{fig:rating_trajectories}, green curve)}. This patterns reflected ongoing uncertainty and continuous integration of evidence. As P13 explained, ``\textit{I kept adjusting my ratings as I read more comments, trying to balance the different perspectives I encountered. I felt that they all made sense.}''

\par Beyond numerical ratings, participants actively re-prioritized attribute importance in response to new information. Attributes initially deemed secondary sometimes rose to primary concern. For instance, P6 initially considered ``Cleanliness'' relatively unimportant, but after reading a comment describing unsanitary conditions, elevated it to their top priority, explaining, ``\textit{I didn't think it mattered much at first, but if the room isn't clean, that would ruin the whole stay for me.}'' Likewise, P9, who initially valued ``Location'' above all, shifted their emphasis to ``Service'' after encountering multiple accounts of indifferent or cold staff, reasoning ``\textit{I realized that even if the location is perfect, poor service could make the trip stressful.}''

\par \textit{\textbf{Complexity of Synthesis, Judgment, and Final Rationales.}} 
% Among the 18 participants, the majority (15/18) expressed satisfaction with the hotel and indicated a willingness to book it.  This outcome aligns with our expectation, given the hotel's above-average satisfaction level within the local context.  We further examined the reasons behind the three participants who chose not to proceed with the booking.  Their justifications varied and were distributed across all three experimental conditions, suggesting no apparent relationship between the binary booking decisions and comment ordering. 
The high rate of participants expressing intent to give a positive final satisfaction evaluation (15/18) was unsurprising, given our selection of a hotel that exceeded average market quality standards. The three participants who reported dissatisfaction were evenly distributed across conditions, with their views shaped by individual preferences rather than the order of comment presentation. This pattern underscores our focus on the cognitive processes at play during evaluation rather than on outcomes alone. Beyond overall satisfaction level, we observed meaningful variation in the confidence and complexity of participants' justifications. Participants in the \textit{Negative-First} and \textit{Interleaved} conditions tended to provide longer, more evidence-based rationales, often citing trade-offs or inconsistencies in the comment content (e.g., ``Good location but noisy nights'' or ``Service seems inconsistent, but price makes up for it''). In contrast, \textit{Positive-First} participants generally offered shorter, sentiment-driven explanations, such as ``Looks like a good deal'' or ``Everyone seems happy.'' These findings indicate that comment presentation order does more than influence initial impressions—it shapes the entire evaluative trajectory, affecting the depth of reasoning and the nature of justification. In other words, the sequencing of information can nudge users toward either quick, affective conclusions or more deliberate, evidence-integrating evaluations.

% The most striking differences appeared in the nature of these justifications. Beyond the binary decision outcome, we observed variation in the confidence and complexity of participants’ justifications. The most striking difference emerged in the nature of participants' final justifications. 

% \par \textit{\textbf{Summary.}}
% We analyzed how participants’ engagement patterns varied across the four decision stages in different comment ordering conditions. As shown in \cref{fig:stage_allocation}, while all participants progressed through the same core stages, the \textit{temporal allocation of attention} differed markedly by conditions.
\subsubsection{Three Critical Issues in Decision-Making Trajectories.}
% From Process to Persona: Three Dominant Decision Modes
% \par \textit{\textbf{Initial Framing Sets a Powerful, Lasting Anchor.}} Our primary quantitative findings establish that the initial framing, dictated by the first few comments, had a decisive and lasting effect on participants' final judgments.

% \par \textit{\textbf{The Iterative Loop: Biased Foraging and Volatile Beliefs.}}
% Analysis of the rating change history logs reveals how these final judgments were formed during the Evidence Foraging and Belief Updating stages. The initial frame did not just set an anchor; it biased the entire evaluation process.

We identified three recurrent issues that shaped participants' decision-making trajectories across the four stages.
\par \textit{\textbf{Issue I: Goal Alienation—From ``Initial Framing'' to ``Verification''}.} The first critical issue arose during Stage 1 (\textit{Initial Framing}), where the initial sequence of comments did more than merely bias perception; it redefined participants' core evaluation goals. Instead of maintaining an open-ended, impartial stance, participants quickly gravitated toward one of two skewed objectives. In the \textit{Positive-First} condition, strong early impressions fostered a tendency toward positive verification, with participants seeking information that reassured and confirmed their initial beliefs rather than conducting a balanced assessment. In contrast, the \textit{Negative-First} condition triggered a risk-defense mindset: early unfavorable cues were perceived as potential threat, prompting participants to focus on mitigating perceived risks rather than holistically weighing the evidence. This early ``goal alienation'' effectively locked participants into a fixed cognitive strategy before they had engaged with the full range of available information, constraining their capacity to revise initial impressions objectively.

\begin{figure*}[h]
    \centering
    \includegraphics[width=\textwidth]{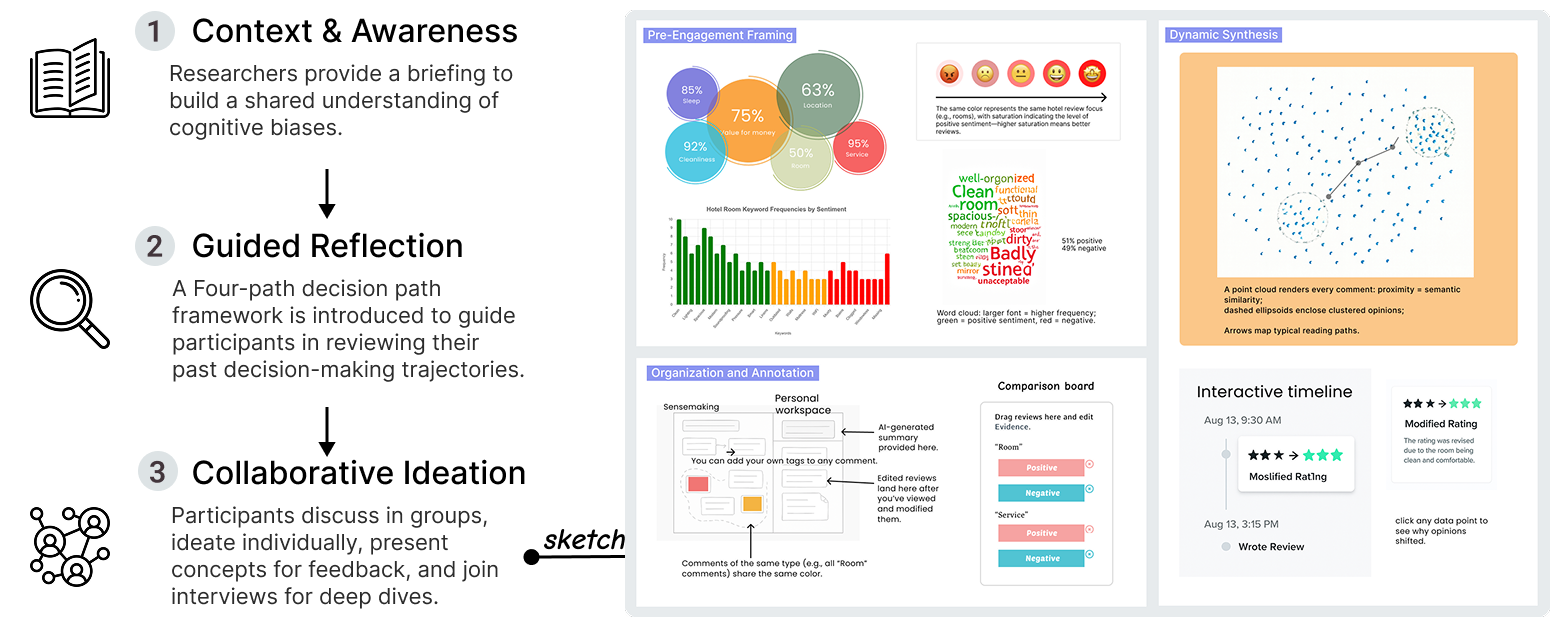}
    \caption{The 45-minute workshop was conducted in three phases: (1) establishing context and raising awareness of cognitive biases, (2) reflecting on decision-making trajectories from Study I, and (3) ideating and developing design concepts in group and individual activities.}
    \label{fig:studyII}
\end{figure*}

\par \textit{\textbf{Issue II: Asymmetric ``Evidence–Belief'' Loops Drive Cognitive Oversimplification or Overload}.} The iterative cycle between Stage 2 (\textit{Evidence Foraging}) and Stage 3 (\textit{Belief Updating}) exhibited two markedly unbalanced behavioral patterns. One group of participants engaged predominantly in confirmatory information seeking with minimal critical scrutiny, leading to premature closure and weak evidential grounding—a form of cognitive oversimplification. In contrast, another group pursued exhaustive searches and frequent belief revisions, creating cognitively taxing loops characterized by sustained uncertainty and mental fatigue. These asymmetries highlight how initial framing can shape the dynamics between evidence gathering and and belief updating, ultimately influencing both the depth of reasoning and the cognitive effort expended.
% We observed that the cyclical process between Stage 2—Evidence Foraging—and Stage 3—Belief Updating diverges into two imbalanced patterns under different conditions. 1) Users tended to confirm their initial impressions without rigorous scrutiny, resulting in a shallow loop characterized by premature closure and insufficient evidential grounding. 2) Users engaged in intensive, compensatory information search and frequent belief revisions, resulting in a laborious loop marked by excessive cognitive effort and uncertainty.

\par \textit{\textbf{Issue III: Insufficient Support for ``Synthesis–Judgment'' Undermines Final Decision Quality}.} During Stage 4 (\textit{Synthesis and Judgment}), we observed that participants frequently forgot key information from earlier comments and had to repeatedly scrolled back to refresh their memory. Lacking interface tools to capture, organize, or compare comment content, this back-and-forth process was inefficient and prone to information loss. Consequently, only participants who engaged in a mentally demanding cycle of intensive rereading and reconstruction were able to reach well-founded conclusions, while most relied on quick heuristics, such as prioritizing the most recent or most extreme comments. These observations suggest that insufficient support for ``synthesis and judgment'' fosters an inefficient, high-effort process that may cause users to rely on cognitive biases, thereby compromising evaluation quality.
% This indicates that final decision quality depended more on individual cognitive effort than on any systemic support for coherent evidence integration.
% 信息longyu

% \section{Study II:  What are the key needs when navigating complex comment environments that may induce biased perceptions?}
% \section{Study II: Identifying key needs when navigating complex comment environments that may induce biased perceptions?}

\section{Study II: Deriving Design Requirements from User-Generated Ideas in a Workshop}
% In Study I, we confirmed that the presentation order of comments significantly induces cognitive bias in users, and confirmed three issues in their during the whole four-phase stage. To address these issues, we conducted Study II. By engaging the same participants in a participatory workshop, we aimed to identify key needs and collaboratively develop interface solutions that could mitigate the biases observed in the initial experiment.

% \ouyang{Our findings in Study I demonstrated that the presentation of comments systematically shapes users' decision-making trajectories, introducing three critical issues across the four stages (\cref{sec:reofstudyI}). This raises our second research question (\textbf{RQ2}): From the user's perspective, what are the key needs when navigating complex comment environments that may induce biased perceptions? To effectively answer this, we conducted Study II, involving the same participants from Study I to leverage their firsthand experience and gain deeper insights.}
\par Our findings from Study I revealed that the way comments are presented systematically shapes users' decision-making trajectories, giving rise to three critical issues across the four stages (\cref{sec:reofstudyI}). These insights motivated a deeper investigation: from the user's perspective, what design requirements are necessary to navigate complex comment environments where biased perceptions may emerge (\textbf{RQ2})? To address this question, we conducted Study II with the same participants from Study I, drawing on their firsthand experience to surface and articulate these requirements in greater depth.

% Through the analysis of user feedback and co-designed sketches, we identified key directions for improving the interface, such as increasing source transparency and incorporating emotional cues. These insights guided our subsequent development of an enhanced interface featuring.
% \subsection{Method}

\subsection{Participants}
\par We invited all 18 graduate students from Study I to participate in the workshop. Their firsthand experience with the task scenario and the challenges they encountered offered valuable insights for collaboratively co-designing effective solutions. 
% \ouyangrese{Study II also obtained IRB approval from the Research Ethics Committee.}
% \ouyangre{We obtained IRB approval from the Research Ethics Committee. All data were anonymized: only participant pseudonyms were retained, and no personally identifiable information was recorded or stored.}

 % Reflective review of decision-making trajectories
\subsection{Workshop Procedure}
% \par The 45-minute participatory design workshop was structured into two main phases to elicit user insights and translate them into actionable design requirements: \textbf{\textit{Phase 1: Context Setting and Reflective Review.}} Unlike the post-task interviews in Study I, which focused primarily on recalling difficulties, this phase encouraged participants to engage in deeper reflection on their decision-making experiences. To guide this process, we introduced an analytical framework distilled from Study I — the four-stage decision path and three critical issues — providing a structured lens through which participants could revisit and critically examine their cognitive journeys. This collective reflection not only validated our prior findings but also enriched and contextualized the problem space, laying a solid foundation for the subsequent design activities. \textbf{\textit{Phase 2: Ideation and development of design concepts.}} In the second half, participants were divided into three groups according to their experimental condition in Study I. Each group began with discussions to align on key issues and share initial thoughts. Participants then individually generated design concepts, followed by two-minute presentations of their ideas to the group, fostering feedback and collective reflection. The workshop concluded with 15-minute semi-structured one-on-one interviews, enabling participants to further articulate their perspectives while also benefiting from exposure to others' approaches and insights throughout the session.

\begin{table*}[h]
\caption{Design concepts (Pre-Engagement Framing, Organization and Annotation, Dynamic Summarization) with their debias strategies, descriptions, and associated design requirements.}
\renewcommand{\arraystretch}{1.3} % 增加行高，让表格更易读
\centering
\resizebox{\textwidth}{!}{%
\begin{tabular}{cccc}
\hline
\textbf{Design Concept} & \textbf{Debias Type} & \textbf{Description} & \textbf{DRs} \\ \hline

\textbf{Pre-Engagement Framing} & Prevent & 
\begin{tabular}[c]{@{}l@{}}Provide an overview of comments and key patterns before engagement, \\ Helping users avoid early anchoring and form balanced initial impressions.\end{tabular} & \textbf{DR1} \\ \hline

\multirow{2}{*}{\textbf{Organization and Annotation}} & Discover & Support collecting, organizing, and annotating comments. & \textbf{DR2} \\ \cline{2-4} 
 & Locate & Help users identify inconsistencies or recurring themes in the workspace. & \textbf{DR3} \\ \hline

\textbf{Dynamic Synthesis} & Mitigate & 
\begin{tabular}[c]{@{}l@{}}Support presenting users’ evolving assessments and \\ Synthesizing their judgments while encouraging reflection.\end{tabular} & \textbf{DR4} \\ \hline

\end{tabular}%
}

\label{tab:design_concepts}
\end{table*}

\par The 45-minute workshop was organized into three phases designed to elicit user insights and translate them into actionable design requirements (see \cref{fig:studyII}). \textit{\textbf{Phase 1: Context establishment and cognitive bias awareness.}} The session began with a researcher-led briefing to establish context. Building on the bias introduction and de-biasing strategies outlined in related work (\cref{sec:bias_re}), this briefing aimed to raise participants' awareness of cognitive biases. The researchers illustrated how interface designs can trigger biases such as anchoring and availability, encouraging participants to engage in more deliberate reflection. \textbf{\textit{Phase 2: Reflective review of decision-making trajectories.}} Next, participants revisited their decision-making experiences from Study I. To guide this process, the analytical framework distilled from Study I—the four-stage decision path and three critical issues (\cref{sec:reofstudyI})—was introduced. This framework provided a structured lens through which participants critically examined their cognitive journeys. The collective reflection not only validated our prior findings but also enriched and contextualized the problem space, laying a solid foundation for subsequent design activities. \textbf{\textit{Phase 3: Ideation and development of design concepts.}} In the final phase, participants were divided into three groups according to their experimental condition in Study I. Each group began with discussions to align on key issues and share initial thoughts. Participants then individually generated design concepts, followed by two-minute presentations to their group, facilitating feedback and collaborative refinement. The workshop concluded with 15-minute semi-structured one-on-one interviews, enabling participants to articulate their perspectives more deeply while also benefiting from exposure to others' approaches and insights throughout the session. Participants were provided US\$5 as compensation for their participation in the workshop.

\subsection{Data Collection and Analysis}
\par We collected all workshop materials, including sticky notes and design sketches. Our analysis focused on characterizing how participants perceived bias and the interventions they proposed to address it. Using affinity diagramming~\cite{lucero2015using}, we organized these ideas into coherent themes, which were then synthesized into concrete design requirements.
% \subsection{Findings: Key Design Goals}
% Our analysis of the workshop data, including participant sketches and discussions, yielded three key design goals. Each goal directly addresses one of the critical issues identified in Study I, translating the lived experiences of our participants into actionable directions for interface design.

% \par \textbf{\textit{DG1: Promote and Maintain an Exploratory Mindset.}}

% \par \textbf{\textit{DG2: Provide Scaffolding for Evidence Foraging and Belief Updating.}}

% \par \textbf{\textit{DG3: Support Efficient Synthesis for an Informed Final Judgment.}}

\subsection{Findings}
% \par \ouyang{Our analysis of the workshop data revealed a rich collection of user-generated design concepts aimed at mitigating specific issues. By synthesizing these concrete ideas, we then derived four overarching design requirements that articulate the core principles.} 
% We first present the emergent concepts, followed by the design goals they inform.

% \usepackage{multirow}
% \usepackage{graphicx}

\subsubsection{User-Generated Design Concepts}
\par Participants sketched a variety of design concepts (see \cref{fig:studyII}). As shown in \cref{tab:design_concepts}, three prominent themes emerged from their creations:
\begin{itemize}
\item \textit{\textbf{Pre-Engagement Framing.}} To mitigate the tendency for users to form snap judgments, several groups suggested adding overview components at the top of the comment section. These components would serve to orient users before they engage with individual remarks. The aim was to highlight broader patterns and key areas of divergence, offering a more balanced starting point. Proposed designs included pie charts showing sentiment distribution and tag clouds summarizing frequently mentioned issues. Additionally, some designs emphasized presenting a spectrum of perspectives upfront—such as differing opinions on service quality or room amenities—to help users approach the comments with an expectation of diversity, rather than relying solely on the first few remarks they encounter.

\item \textit{\textbf{Sensemaking through Organization and Annotation.}} To help manage the cognitive load associated with processing a large volume of comments, the most common suggestion was a ``personal workspace'' or ``comparison board''. Participants envisioned a persistent sidebar where users could drag and drop noteworthy comments and annotate them with personal notes. This feature allowed users to externalize their thoughts and maintain context while navigating through the comment section. Some design proposals included the use of ``visual clusters'' to group recurring themes, which facilitated reflection and enabled users to make more deliberate judgments.

\item \textit{\textbf{Synthesis through Dynamic Summarization.}} For the final evaluation process, participants suggested the inclusion of a ``Final Report'' or ``Judgment Summary'' page. This page would allow users to track changes in their assessments over time, providing a clear view of how their opinions evolved. Some designs also incorporated features that prompted users to provide brief justifications for any significant shifts in their judgments, encouraging deeper engagement with the evaluation process.

% The goal of these concepts was to prime users for complexity from the outset.
% \item An Interactive Workspace for Sensemaking. To manage the cognitive load of processing numerous comments (Issue II), the most frequent concept was a "personal workspace" or "comparison board." Participants envisioned a persistent sidebar where they could drag-and-drop noteworthy comments. Their sketches included features for annotating these saved comments with personal notes, applying color-coded tags (e.g., `red` for negative, `green` for positive), and filtering their collection by these tags. 
% % This demonstrated a clear desire to externalize the process of sorting and comparing evidence.
% \item \textit{\textbf{A Dynamic Summary for Final Synthesis.}} To aid the final decision-making process (Issue III), participants conceived of a "Final Report" or "Decision Summary" page. This was often depicted as a printable or saveable view that automatically aggregated all the comments and notes from their workspace. Some designs included a feature that would place the user's initial and final ratings side-by-side, prompting them to write a short rationale for any significant changes in their judgment. 
% This pointed to a need for a structured tool to facilitate a final, evidence-based conclusion.
\end{itemize}

\begin{table*}[h]
\centering
\caption{\ouyangre{Number of design concepts generated by each group across three types.}}
\label{tab:concept_counts}
\begin{tabular}{lcccc}
\hline
Group & Pre-Engagement Framing & Organization \& Annotation & Dynamic Summarization & Total \\
\hline
Positive-First & 6 & 3 & 1 & 10 \\
Negative-First & 3 & 5 & 3 & 11 \\
Interleaved & 2 & 4 & 4 & 10 \\
\hline
\end{tabular}
\end{table*}

\par \ouyangre{Notably, participants’ prior exposure to specific conditions in Study I appeared to influence the aspects of the workshop on which they focused. As shown in \cref{tab:concept_counts}, the Positive-First group generated 6 concepts related to Pre-Engagement Framing, 3 for Organization and Annotation, and 1 for Dynamic Summarization, highlighting a focus on orienting users, emphasizing key patterns, and mitigating early anchoring. The Negative-First group contributed 3 concepts for Pre-Engagement Framing, 5 for Organization and Annotation, and 3 for Dynamic Summarization, reflecting a stronger emphasis on structuring information and supporting reflective review. The Interleaved group produced 2 concepts for Pre-Engagement Framing, 4 for Organization and Annotation, and 4 for Dynamic Summarization, suggesting a balanced approach that integrates organization and synthesis to support comprehensive evaluation. These patterns are descriptive rather than conclusive due to the small group sizes ($N = 6$ per group). Future efforts with larger samples could elucidate how initial conditions shape participants’ focus on different types of design concepts.}

\subsubsection{Key Design Requirements}
\par To inform the design of our tool, we synthesized the user-generated concepts and distilled them into four core design requirements:

\par \textbf{DR1: Encouraging and Maintaining an Exploratory Mindset.} To mitigate the effects of early anchoring and goal-alienation, the probe should encourage an open, curious approach from the beginning. This involves framing the comment space in a way that primes users to expect diverse perspectives and unresolved trade-offs, rather than merely seeking confirmation. Pre-engagement elements such as summary visualizations or highlights of controversial topics can help guide users toward broader exploration and prevent premature conclusions.

\par \textbf{DR2: Providing Interactive Supports for Organizing and Annotating Comment Evidence.} To mitigate cognitive overload and prevent unbalanced evidence-belief loops, the probe should offer lightweight, persistent scaffolds for sensemaking. By enabling users to collect and organize key comment elements—through actions such as tagging—they can externalize their thought processes, track their evolving impressions, and engage more thoughtfully with the content.

\par \textbf{DR3: Facilitating In-Situ Reflection for Enhanced Critical Assessment.} To interrupt impulsive judgment and reduce the risk of biased closure, the probe should incorporate subtle, well-timed cues that encourage users to pause and reassess their developing conclusions. These prompts—triggered by instances of rapid decision-making, prolonged indecision, or sharp shifts in ratings—offer users an opportunity to revisit earlier evidence, challenge conflicting viewpoints, and refine their reasoning before progressing. Instead of enforcing reflection, these interventions gently prompt reconsideration at critical moments.

\par \textbf{DR4: Supporting Dynamic Synthesis through Integrated Interpretation.} Final judgments are often hindered by the difficulty of consolidating fragmented impressions accumulated over time. The probe should assist with retrospective synthesis by offering dynamic overviews of saved content, visual contrasts between early impressions and final decisions, and optional guidance for articulating underlying reasoning. Such support can help identify internal inconsistencies, encourage reflective thinking, and bolster confidence in the final conclusion.

\section{CommSense}
\par
\begin{figure*}[h]
    \centering
    \includegraphics[width=1\linewidth]{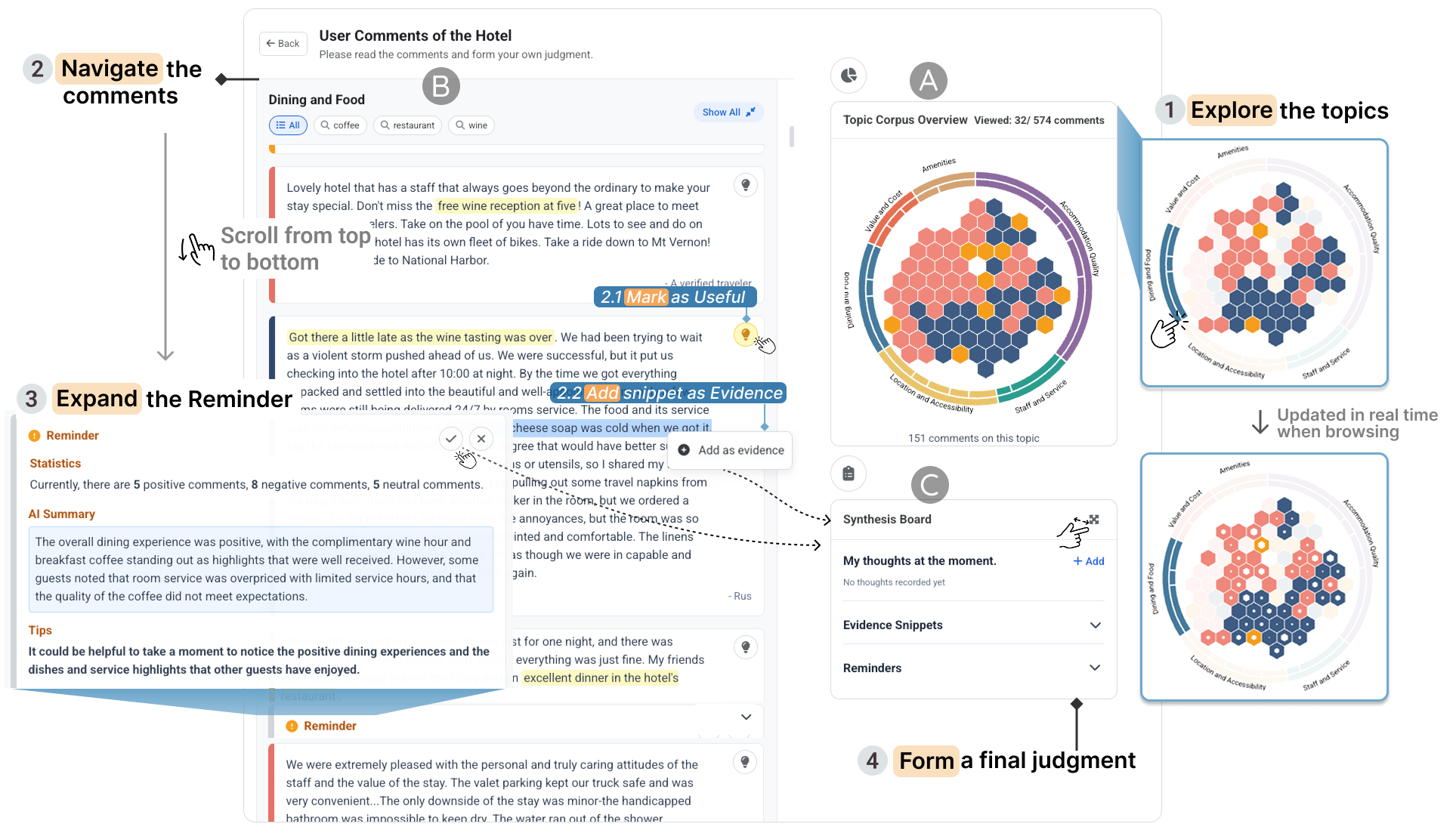}
    \caption{Our interface comprises three parts: (A) the \textit{Topic Corpus Overview}, (B) the \textit{Comment Navigation Panel}, and (C) the \textit{Synthesis Board}. The \textit{Topic Corpus Overview} and \textit{Synthesis Board} are integrated into a side toolbar positioned to the right of the \textit{Comment Navigation Panel}.
    The workflow guides users to: (1) explore the topic, (2) navigate comments while adding evidence snippets and marking relevant ones, (3) expand the reminder, and (4) form a final judgment.    
    }
    \label{fig:final_interface}
\end{figure*}

 Based on the design requirements identified in our empirical studies, we developed \textit{CommSense} (\cref{fig:final_interface}), a lightweight tool that helps users engage with online comments more rationally and reflectively, mitigating biased impressions and supporting more balanced judgments.

% We introduce a lightweight reminder strategy in comment navigation. When users repeatedly consume comments of the same polarity, the interface surfaces a subtle cue (e.g., “You have mainly read positive opinions—consider checking the other side”) to encourage balance. Additionally, after sufficient browsing, another reminder (e.g., “You have gathered enough evidence—would you like to summarize now?”) gently prompts transition toward synthesis. Both cues are designed to appear inline while scrolling, remaining non-intrusive yet timely.

\subsection{Interface Design}
\par As shown in \cref{fig:final_interface}, the interface comprises three coordinated parts: the \textit{Topic Corpus Overview} (\cref{fig:final_interface}-A), the \textit{Comment Navigation Panel} (\cref{fig:final_interface}-B), and the \textit{Synthesis Board} (\cref{fig:final_interface}-C). The \textit{Topic Corpus Overview} and the \textit{Synthesis Board} are integrated into a side toolbar positioned to the right of the \textit{Comment Navigation Panel}.

\subsubsection{Topic Corpus Overview}
\par The \textit{Topic Corpus Overview} serves as a concise entry point into the comment corpus, helping users quickly grasp how opinions are distributed across key topics \textbf{(DR1)}. At the center of this view is a hexbin plot, where comments are embedded in a reduced-dimensional semantic space and clustered by thematic similarity. These clusters are aggregated into hexagonal bins, each containing up to 15 items—a threshold chosen to balance information density with readability. Sentiment polarity is encoded with a trichromatic scheme (red-orange for negative, dark blue for positive, and orange for neutral), providing an immediate visual cue of opinion trends. Additional details on the backend processing pipeline are provided in \cref{sec:imdetails}.
% The design offers an intuitive entry point for information foraging, encouraging further exploration of the comments.
\par Surrounding the hexbin are two interactive rings: an \textbf{outer ring} that represents broad thematic areas (e.g., ``Staff and Service'' or ``Location and Accessibility''), and an \textbf{inner ring} that displays the specific keywords within those areas (e.g., ``staff''). Selecting elements from either ring dynamically updates the hexbin plot and filters the detailed comments in the \textit{Comment Navigation Panel}, automatically expanding the corresponding topic for deeper exploration.
\par As users browse and scroll through comments in the \textit{Comment Navigation Panel}, the plugin tracks which items have been viewed. This information is reflected in the overview, where hexagons associated with the viewed comments are progressively masked, providing immediate visual feedback on the portions of the semantic space already explored. A real-time progress counter beneath the visualization further indicates the number of viewed comments relative to the total selected.
\par Together, these features support intuitive and structured exploration: users can maintain awareness of the overall comment landscape, drill down into topics of interest, and monitor their progress in real time, encouraging more reflective and well-informed engagement with the corpus.

% Users can interact with the overview to select specific thematic areas—such as “Room,” “Service,” or “Location”—which then guides subsequent, topic-focused exploration in the left \textit{Comment Navigation Panel}. As users browse comments in \textit{Comment Navigation Panel}, individual hexagons corresponding to the viewed comments are shaded, providing immediate visual feedback on which areas have already been explored. Additionally, the real-time distributional statistics of viewed comments across topics are presented below the figure.

\subsubsection{Comment Navigation Panel}
\par The \textit{Comment Navigation Panel} serves as the primary workspace, enabling users to browse, filter, and engage with comments (\textbf{DR2, DR3}).

\par \textbf{Organization Structure.} To foster balanced, evidence-based impressions, the system organizes comments by emphasizing contrasting viewpoints rather than relying solely on algorithmic ranking. The design explicitly prioritizes both diversity and polarity of opinions. Within each thematic category (e.g., Room or Location, derived from the \textit{Topic Corpus Overview}), users are first presented with a pair of semantically similar comments expressing opposing sentiments (\textbf{DR1}), immediately highlighting the spectrum of perspectives. This approach not only surfaces diverse opinions but also encourages consideration of counter-evidence that might otherwise be overlooked. Users can further tailor their exploration by selecting keywords, with the Overview dynamically updating to reflect their choices. Moreover, as users scroll, the system applies semantic analysis to proactively recommend contextually relevant comments with opposite sentiments, ensuring ongoing exposure to a balanced distribution of viewpoints.

% Additionally, the upcoming five comments in the scroll path are highlighted in the overview panel, providing a continuous link and a short-term preview between the detailed view and the high-level summary.
% To do
% [pos-neg 正负各一条组]

% This design allows users to navigate the information space efficiently, locate relevant evidence, and provides the foundational structure upon which our intervention mechanisms operate.
\par \textbf{Active Annotation.} The \textit{Comment Navigation Panel} supports two types of explicit annotation \textbf{(DR2)}. First, users can mark entire comments as ``useful'', providing a lightweight indication of relevance. Second, they can engage more deeply by highlighting specific text fragments within comments and saving them as ``Evidence Snippets''. Each snippet is enriched with contextual metadata, such as topic and sentiment, and automatically organized within the \textit{Synthesis Board}.

\par \textbf{Lightweight Reminder.} The lightweight reminder is designed to foster sensemaking and in-situ reflection during comment navigation \textbf{(DR3)}. Delivered inline as subtle, dismissible cues, it helps direct user attention and encourage balanced engagement with the comment set. The strategy integrates two complementary interventions that address common challenges in \textit{Evidence Foraging and Belief Updating}:
\par \textit{1) Encouraging Viewpoint Diversity.} To counter selective exposure, the tool tracks the polarity of comments marked as ``useful'' and compares it against the overall distribution of comment valences. When a user's selections diverges significantly from the global distribution (e.g., mostly positive or negative), a cue appears (e.g., ``You have mainly read negative opinions—consider checking the other side''), nudging exploration of underrepresented perspectives.
\par \textit{2) Promoting Exploration Across Topics.} To support the shift from evidence gathering to higher-order judgment, the tool monitors browsing coverage, i.e., the proportion of comments viewed within each topic. Once users have examined a substantial share of comments for the current topic (e.g., 70\%), a one-time cue appears (e.g., ``You have explored 70\% comments on this topic—take a moment to summarize key insights and explore other topics, such as `Dining and Food'''). This encourages consolidation of insights while prompting exploration of overlooked areas.
\par The Reminder is presented as a multi-part artifact seamlessly integrated into the comment stream, salient yet non-intrusive. Each reminder comprises three elements: \textit{Statistical Grounding}, which displays the metrics that triggered the cue (e.g., counts of positive vs. negative comments, number of ``useful'' items), anchoring the reminder in evidence; \textit{AI Summary}, which provides a concise, dynamically generated overview of recently viewed comments alongside an actionable suggestions; and \textit{User Mind}, a dedicated text field for users to externalize reflections. Users can choose to save a reminder to the persistent \textit{Synthesis Board} by selecting `Add', or 'Dismiss' it to continue browsing.

\subsubsection{Synthesis Board}
\par The \textit{Synthesis Board} supports the final synthesis (\textbf{DR4}) by integrating three modalities: (1) \textit{User's Instant Thoughts}, capturing spontaneous, user-authored insights; (2) \textit{Evidence Snippets}, drawn from annotated comment excerpts; and (3) \textit{Reminders}, selectively saved during navigation. By collocating these artifacts within a unified, expandable space, the board enables users to consolidate insights and articulate well-grounded final judgments.

\begin{figure*}[h]
    \centering
    \includegraphics[width=\linewidth]{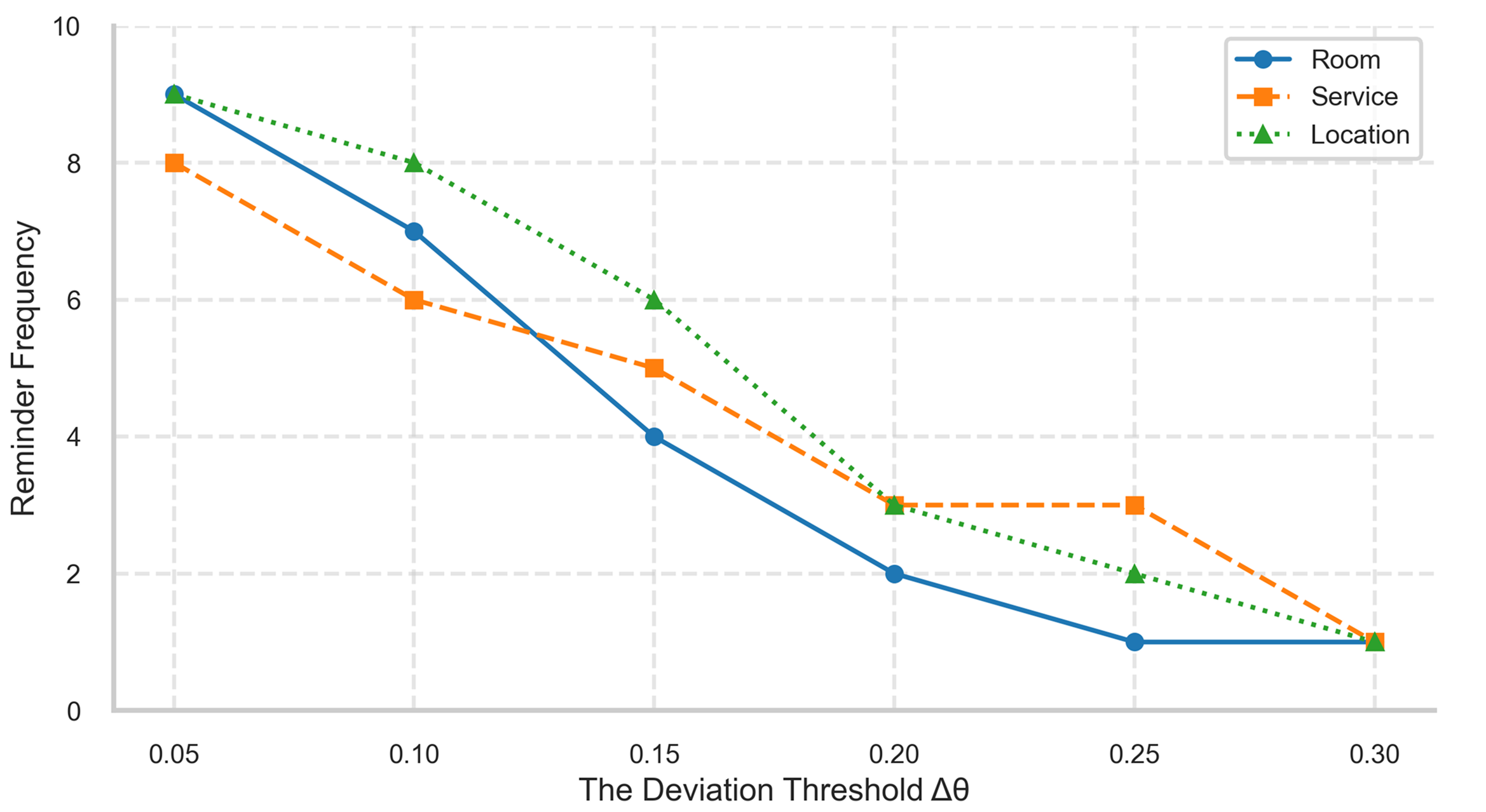}
         \caption{\ouyangre{Relationship between sentiment deviation threshold ($\Delta \theta$) and reminder frequency across topic comments.}}
    \label{fig:delta_theta_testing}
\end{figure*}

\subsection{Implementation Details}
\label{sec:imdetails}
\par \textit{CommonSense} is implemented using modern web technologies, featuring a Vue.js\footnote{\url{https://vuejs.org/}} frontend and a Flask\footnote{\url{https://flask.palletsprojects.com/}} backend that provides API services. The system is deployed on a web server, allowing users to access it conveniently through standard web browsers. The dataset used is described in \cref{sec:dataset}.

% \label{sec:ce}
% \subsubsection{Data}
% @ gaoshh1
\subsubsection{Data Progressing Pipeline}
\label{sec:datapipeline}

\par We integrate advanced methodologies to address the complexity of online comments, which often span multiple dimensions and perspectives. Using a structured processing pipeline, our tool extracts meaningful insights from these diverse user expressions.

\par \textbf{Text Embedding and Dimensionality Reduction.} To capture nuanced text semantics, we encode each comment into a high-dimensional vector using the ``paraphrase-multilingual-mpnet-base-v2''~\cite{reimers-2019-sentence-bert} model. To represent semantic differences between comments, we apply nonlinear dimensionality reduction technique UMAP~\cite{mcinnes2018umap}, a well-established technique known for its effectiveness in preserving the local and global cluster structure of high-dimensional data~\cite{jeon2023classes}.

\par \textbf{Topic and Keyword Extraction.} Keywords were extracted from comments using KeyBERT~\cite{grootendorst2020keybert}, enhanced with the same model used for text embeddings, 'paraphrase-multilingual-mpnet-base-v2', to ensure consistency. The 20 most frequent keywords were then summarized and categorized into six distinct topics using an LLM\footnote{We use GPT-4o as the LLM model. The prompts are provided in \cref{sec:props_top_ex}.}. Each keyword was assigned to one of these topics. To facilitate efficient content retrieval, the LLM was further employed to extract comment segments relevant to each keyword, enabling users to quickly locate pertinent information.

\par \textbf{Sentiment analysis.} In addition to semantic and topical structures, we also analyze the emotional polarity of comments. Using \textit{multilingual-sentiment-analysis}~\cite{loureiro2022timelms}, each comment is classified as positive, negative, or neutral. When focusing on specific topics, these classifications can be aggregated to show the sentiment distribution within each topic. 
% \ouyangre{We opted for this approach over an LLM, as the sentiment analysis task in this study constitutes a well-defined three-class problem that established models can execute efficiently and reliably. This choice enables rapid processing of large volumes of comments while maintaining consistent and reproducible results, avoiding the higher computational costs and potential variability associated with LLM outputs.}

\par By capturing the semantic structure, topical organization, and sentiment of online comments, these approaches collectively constitute the preprocessing stage of our tool.

% Qwen3-235B-A22B-Instruct-2507

\begin{table*}[h]
\centering
\caption{\ouyangre{Results of pilot testing for topic view coverage thresholds across three topics (Room: 40 comments, Service: 60 comments, Location: 70 comments). \textit{Comments before reminder} shows the number of comments when the reminder was triggered, and \textit{Comments after reminder} shows the median number of additional comments viewed after the reminder.}}
\label{tab:view_coverage_testing}
\resizebox {\linewidth} {!}{
\begin{tabular}{>{\hspace{0pt}}m{0.09\linewidth}>{\centering\hspace{0pt}}m{0.165\linewidth}>{\centering\hspace{0pt}}m{0.165\linewidth}>{\centering\hspace{0pt}}m{0.165\linewidth}>{\hspace{0pt}}m{0.348\linewidth}} 
\toprule
\textbf{Coverage }\par{}\textbf{Threshold} & \textbf{Room }\par{}\textbf{(Comments Before / After)} & \textbf{Service }\par{}\textbf{(Comments Before / After)} & \textbf{Location }\par{}\textbf{(Comments Before / After)} & \textbf{Assessment} \\ 
\midrule
50\% & 20 / 9 & 30 / 7 & 35 / 8 & Too early; users had not yet formed an understanding of the topic \\
60\% & 24 / 7 & 36 / 6 & 42 / 6 & Slightly early; some users still not ready \\
\textbf{70\%} & 28 / 3 & 42 / 3 & 49 / 3 & \textbf{Balanced; chosen threshold} \\
80\% & 32 / 2 & 48 / 1 & 56 / 2 & Slightly late; some reminders triggered too late \\
90\% & 36 / 1 & 54 / 1 & 63 / 2 & Too late; reminders rarely triggered \\
\bottomrule
\end{tabular}
}
\end{table*}

\subsubsection{Reminder Trigger Mechanism}
\par The reminder strategy consists of two complementary interventions. First, to ensure balanced sentiment sampling, a reactive state array tracks comments marked as ``useful.'' After each update, the local sentiment distribution is recomputed and compared to the topic's global corpus distribution obtained at initialization. For each topic $t$, an intervention is activated if the deviation exceeds a predefined threshold, formally expressed as:
\begin{equation}
\Delta p(t) = 
\max \Big(
\big|p_{\text{local}}^{\text{neg}}(t) - p_{\text{global}}^{\text{neg}}(t)\big|,\;
\big|p_{\text{local}}^{\text{pos}}(t) - p_{\text{global}}^{\text{pos}}(t)\big|
\Big) 
> \Delta\theta,
\end{equation}

where  

\[
\begin{aligned}
p_{\text{local}}^{\text{neg}}(t) &= \frac{|S_{\text{useful},t}^{\text{neg}}|}{N_{\text{useful},t}}, &
p_{\text{global}}^{\text{neg}}(t) &= \frac{|S_{\text{corpus},t}^{\text{neg}}|}{N_{\text{corpus},t}},\\
p_{\text{local}}^{\text{pos}}(t) &= \frac{|S_{\text{useful},t}^{\text{pos}}|}{N_{\text{useful},t}}, &
p_{\text{global}}^{\text{pos}}(t) &= \frac{|S_{\text{corpus},t}^{\text{pos}}|}{N_{\text{corpus},t}}.
\end{aligned}
\]

\par Here, $S_{\text{useful},t}^{\text{neg}}$ and $S_{\text{useful},t}^{\text{pos}}$ denote the subsets of \textit{useful} comments with negative and positive sentiment, respectively, in topic $t$, and $N_{\text{useful},t}$ is the total number of useful comments in that topic. Similarly, $S_{\text{corpus},t}^{\text{neg}}$ and $S_{\text{corpus},t}^{\text{pos}}$ represent all negative and positive comments in the corpus for topic $t$, with $N_{\text{corpus},t}$ as the total number of corpus comments. 
% $\Delta\theta$ is a tunable deviation threshold (set empirically, e.g., 0.20), determined through internal testing to maintain a balanced perceived sentiment during user evaluation. 

\ouyangre{The deviation threshold $\Delta\theta$ was established as a tunable design parameter and set to 0.20 to support balanced perceived sentiment throughout the user experience. This value was determined through iterative pilot testing (see \cref{fig:delta_theta_testing}), which monitored reminders across three topics (Room, Service, and Location), each with 40 comments. Thresholds below 0.15 elicited excessively frequent prompts, while thresholds above 0.25 delayed interventions, diminishing their effectiveness.
Selecting $\Delta\theta = 0.20$ provided a consistent balance, yielding reminders that were both timely and unobtrusive. Notably, these thresholds were determined in a small-scale pilot test, rather than through precise experimental measurement. The formula is evaluated only when $N_{\text{useful},t} > 0$ to prevent division-by-zero errors and mitigate small-sample noise.}

\par Second, to encourage broader topic exploration, the tool non-blockingly tracks which comments enter the viewport. For each topic, a \texttt{Set} stores the unique IDs of viewed comments. 
% When 70\% of a topic’s comments have been viewed, a reminder is triggered. A boolean flag ensures the reminder for each topic is dispatched only once to prevent redundancy. 
\ouyangre{When a certain proportion of a topic’s comments have been viewed, a reminder is triggered, with a boolean flag ensuring that each topic triggers the reminder only once to prevent redundancy. To identify an appropriate coverage threshold, we conducted pilot testing with 5 participants. Each participant independently browsed all three topics—Room (40 comments), Service (60 comments), and Location (70 comments). As summarized in Table~\ref{tab:view_coverage_testing}, a 50\% threshold triggered reminders prematurely (e.g., after users had viewed only 20 comments for Room), whereas 80\% or 90\% thresholds delayed reminders until most comments had already been browsed. Based on these observations, we selected a 70\% coverage threshold as the final setting, which provided a balanced intervention: reminders occurred after users had viewed the majority of comments (median), effectively encouraging further exploration without being intrusive.}

\par Upon activation of either trigger, the reminder content is dynamically assembled. The \textit{Statistical Grounding} component is populated directly from the computed metrics, while the GPT-4 API is queried to generate a concise AI Summary accompanied by an actionable suggestion. The prompts used are provided in \cref{sec:props_AI_summary}.

% \subsubsection{AI Summary.}

% \vspace{-1mm}

% \section{Evaluation}
\section{\textbf{Study III}: To What Extent can Interface-Level Strategies Reduce Bias Induced by Comment Presentation?}
\par To investigate whether \textit{CommSense} mitigates biased impressions, increases bias awareness, and promotes reflective judgment \textbf{(RQ3)}, we conducted a between-subjects study, comparing \textit{CommSense} with a baseline interface resembling online comment sections.

\subsection{User Study Design}
\subsubsection{Baseline and Data Settings}
\par
\begin{figure*}
    \centering
    \includegraphics[width=\linewidth]{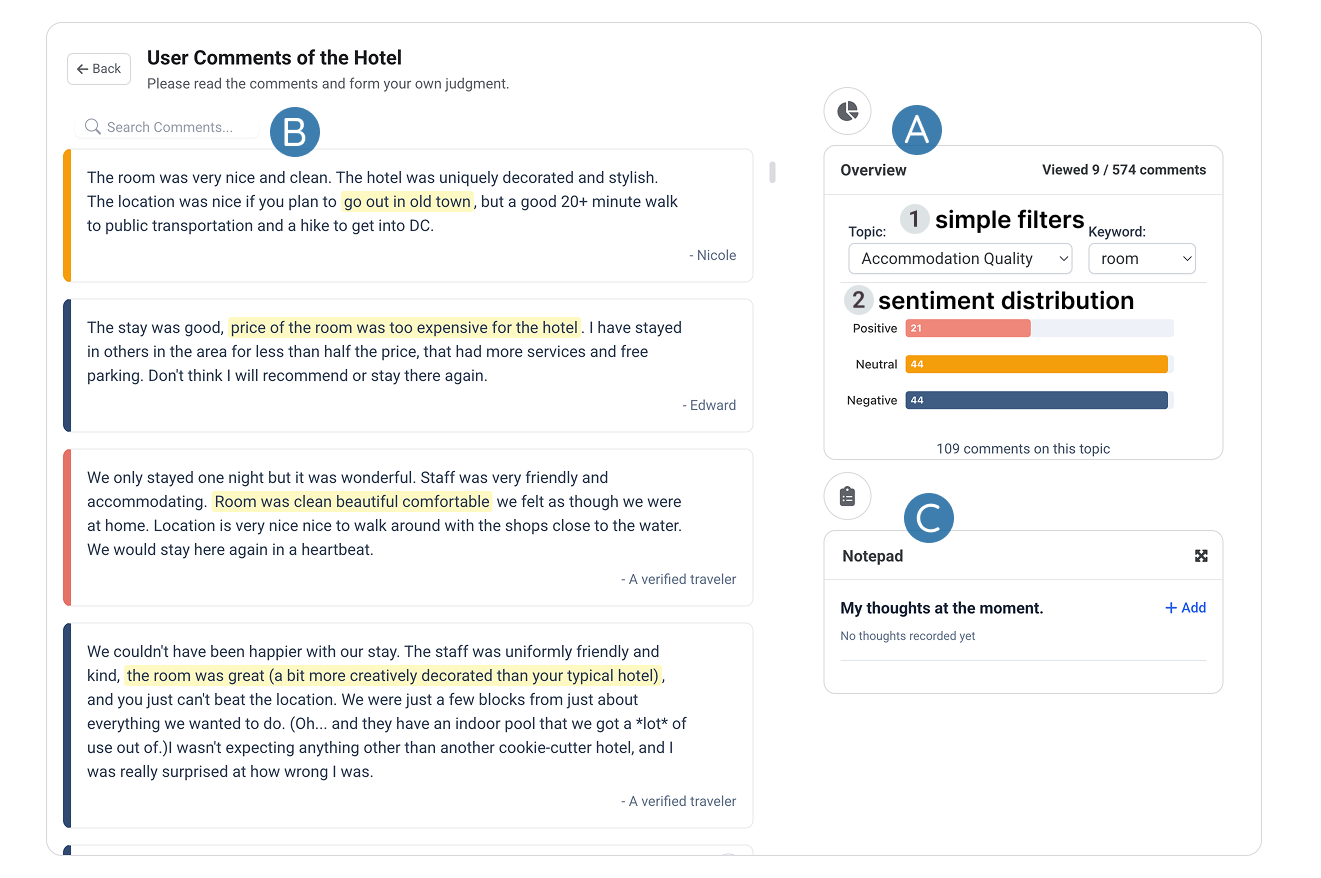}
    % \vspace{-4mm}
    \caption{\ouyangre{
    % The baseline interface for our user study includes three components: (A) the \textit{Comment Overview}, (B) the \textit{Comment Navigation Panel}, and (C) the \textit{Notepad}.
    The baseline interface for our user study includes three components: (A) the \textit{Comment Overview}, which integrates two basic features, allowing users to apply (1) simple filters and view the (2) sentiment distribution; (B) the \textit{Comment Navigation Panel}, where users can browse and access comment details; and (C) the \textit{Notepad}, where users can record their thoughts.
    }}
    \label{fig:baseline}
\end{figure*}

 The baseline interface retained a panel visually consistent with \textit{CommSense}'s \textit{Comment Navigation Panel} (see \cref{fig:final_interface}-B) but was intentionally simplified \ouyangre{(see \cref{fig:baseline})}. Specifically, the \textit{Topic Corpus Overview} was replaced with six categorical filters defined by associated keywords, and the \textit{Synthesis Board} was substituted with a plain, unformatted notepad. This design closely resembled conventional comment sections on mainstream platforms, chosen for its prevalence, simplicity, and user familiarity. \ouyangre{Regarding the workflow, users could apply filters and view sentiment distributions (\cref{fig:baseline}-A), browse individual comments in detail (\cref{fig:baseline}-B), and simultaneously record their thoughts and observations in the notepad (\cref{fig:baseline}-C). This concurrent, minimal-support workflow reflected the typical structure of mainstream online comment sections and contrasted with \textit{CommSense}.}

% It thus served as an appropriate reference point for evaluating how \textit{CommSense}'s design strategies shaped comment navigation.

\par To control for information exposure and maintain consistency across conditions, both interfaces accessed the same database described in \cref{sec:dataset}, consisting of 574 real comments collected over the previous six months. The data were processed and delivered through an identical backend pipeline (see \cref{sec:datapipeline}).

\subsubsection{Participants}
\par We recruited 24 participants (12 female, 12 male; aged 20–28) from a local university through online bulletin boards and mailing lists. All participants had prior experience using online travel platforms and reported regularly consulting user comments when evaluating hotels. Participants were randomly assigned to one of two conditions: the baseline interface group (N=12) or the \textit{CommSense} group (N=12). 
% A between-subjects design was employed to avoid carryover effects, such as heightened bias sensitivity after exposure to \textit{CommSense}. Each participant engaged with only one interface condition. 
A between-subjects design was employed, with each participant engaging with only one interface condition. All received \$15 compensation. 
% \ouyangrese{Study III also obtained IRB approval from the Research Ethics Committee.} 
None had taken part in Studies I or II, ensuring independence across samples. Their information are shown in \cref{sec:ParticipantsinstudyIII}. 

\subsubsection{Measurements}
\label{sec:evadims}
\par We adopted three primary dimensions for quantitative evaluation: \textbf{\textit{Perception}}, \textbf{\textit{Functionalities}}, and \textbf{\textit{Final Judgment Assessment}}.

\par The \textbf{\textit{Perception}} dimension captured participants' subjective experiences, focusing on usability and cognitive workload. We employed the System Usability Scale (SUS)~\cite{bangor2008empirical} and NASA-TLX~\cite{hart1988development}, both widely used and validated instruments, to assess participants‘ impressions of system usability and cognitive workload.

\par The \textbf{\textit{Functionalities}} dimension assessed how participants interacted with and leveraged the system's core components. First, participants completed a \textit{Functionality Assessment}, which evaluated key capabilities: \textit{Overview Providing} \textbf{(DR1)} in the \textit{Topic Corpus Overview}, offering summaries and highlighting key patterns to support balanced impressions; \textit{Comment Organizing} \textbf{(DR2)} in the \textit{Comment Navigation Panel}, enabling structured management of comments; \textit{Evidence Collecting} \textbf{(DR2)} in the same panel, supporting annotation of specific snippets; \textit{Reminder Triggering} \textbf{(DR3)} across the workspace, highlighting inconsistencies or recurring themes; and \textit{Assessment Synthesizing} \textbf{(DR4)} in the \textit{Synthesis Board}, facilitating the consolidation of evolving judgments. These metrics were specifically evaluated by the \textit{CommSense} group. Second, \textit{Interaction Behaviors} were analyzed based on participants' usage logs across the two systems. This included engagement with the \textit{Overview (Filter)} view, capturing interaction frequency and duration; interactions within the \textit{Comment Navigation} panel, capturing semantic distribution (topic spread) and polarity distribution (sentiment balance); and engagement with the \textit{Board (Note)} view, recording usage duration, number of interactions, and notes created.
% These indicators allowed us to evaluate the practical utility of each feature in supporting information exploration, navigation, and reflective engagement.

\begin{table*}[h]
\centering
\caption{The statistical user feedback with \textit{Baseline} and \textit{CommSense}. $p$-values: * $p<.050$, ** $p<.010$, *** $p<.001$.}
\label{tab:commense_perception}
\begin{tabular}{lccccc} 
\hline
\textbf{Categories} & \textbf{Factors} & \begin{tabular}[c]{@{}c@{}}\textbf{Baseline }\\\textbf{Mean/S.D.}\end{tabular} & \begin{tabular}[c]{@{}c@{}}\textbf{\textit{CommSense} }\\\textbf{Mean/S.D.}\end{tabular} & \textbf{\textbf{\textit{P}-value~}} & \textbf{Sig.} \\ 
\hline
\multirow{6}{*}{System Usability Scale} & Easy to use & 5.00 / 1.41 & 5.50 / 1.09 & 0.435 &  \\
 & Functions & 4.5 / 1.24 & 6.17 / 0.72 & 0.0014 & ** \\
 & Quick to learn & 5.83 / 0.94 & 6.00 / 0.74 & 0.602 &  \\
 & Frequency & 3.67 / 1.15 & 5.50 / 1.00 & 0.0015 & ** \\
 & Confidence & 5.17 / 1.34 & 6.17 / 0.83 & 0.082 &  \\
 & Inconsistency & 2.75 / 1.22 & 2.17 / 1.34 & 0.186 &  \\ 
\hline
\multirow{6}{*}{NASA Task Load Index} & Mental Demand & 4.58 / 1.00 & 3.17 / 1.27 & 0.008 & ** \\
 & Physical Demand & 2.92 / 1.16 & 2.00 / 0.74 & 0.041 & * \\
 & Temporal Demand & 3.50 / 0.90 & 2.58 / 1.08 & 0.046 & * \\
 & Performance & 5.17 / 1.27 & 6.00 / 0.95 & 0.108 &  \\
 & Effort & 3.50 / 0.90 & 2.58 / 1.08 & 0.858 &  \\
 & Frustration & 1.83 / 0.58 & 1.58 / 0.51 & 0.307 &  \\
\hline
\end{tabular}
\end{table*}

\par The \textbf{\textit{Final Judgment Assessment}} dimension evaluated the quality of participants' final judgments. Metrics included: \textit{logical consistency}, assessing coherence and reasoning integrity; \textit{completeness}, capturing the extent to which multiple perspectives were considered; \textit{reflective depth}, examining engagement with diverse viewpoints and self-reflection; and \textit{evidence support}, evaluating whether judgments were grounded in concrete comments snippets.

\subsubsection{Study Setup and Procedure}
\par We first conducted a pilot study with two members of the research team (one per condition) to validate task instructions, timing, and evaluation measures. Based on their feedback, we refined instructional materials and adjusted task duration to ensure balance across conditions. The main user study consisted of four stages: \textit{\textbf{Introduction}}, \textit{\textbf{Comment Navigation Task}}, \textit{\textbf{Evaluation and Peer Scoring}}, and \textit{\textbf{Feedback}}.

\par The study began with the \textit{\textbf{Introduction (10 minutes)}}, during which the researcher explained the overall procedure, including the sequence of activities and their expected duration. Participants were introduced to the research context and asked to imagine themselves preparing to evaluate a single hotel for an upcoming academic conference in an unfamiliar city. The task emphasized forming a comprehensive judgment of one hotel, rather than comparing multiple options. Using the instructional materials, the researcher explained the workflow and the assigned interface, ensuring participants clearly understood the task objectives. Any questions were addressed prior to beginning the main task.

\par Next, in the \textit{\textbf{Comment Navigation Task (25 minutes)}}, participants completed the main task using their assigned interface (baseline or \textit{CommSense}). They were asked to freely explore the hotel comments and, within the allotted time, formulate an overall evaluation of the hotel. The system logged participants' interactions and task completion times for later analysis. Unlike Study I, which adopted a more controlled design to isolate specific effects, this study simulated a realistic decision-making context by providing 574 real comments. Participants were free to proceed with their evaluation once they felt they had reviewed a sufficient number of comments.
% These justifications were anonymized and redistributed for peer scoring, either within or across groups, using a rubric adapted from concept-map evaluations, emphasizing clarity, comprehensiveness, and logical structure. This procedure ensured that peer scoring did not lead to bias leakage between conditions while providing a measure of the quality and reflective depth of participants’ judgments.

\par Following this, during the \textit{\textbf{Evaluation and Peer Scoring (20 minutes)}}, participants completed a post-task questionnaire, which included the System Usability Scale (SUS) and NASA-TLX. Participants in the \textit{CommSense} condition additionally evaluated the system's functionalities as implemented in the \textit{Topic Corpus Overview}, \textit{Comment Navigation Panel}, and \textit{Synthesis Board}. Further, once all participants completed the main task, their final judgments were anonymized and redistributed for peer evaluation. Each participant reviewed the judgments submitted by the other 23 participants, ensuring that every submission received multiple independent assessments.

\par Finally, in the \textit{\textbf{Feedback}} stage \textbf{(15 minutes)}, semi-structured interviews were conducted with each participant to elicit deeper reflections on their experience. Participants first explained their reasoning processes during the task and provided overall impressions of the interface. Those in the baseline condition reflected on their experience with the conventional feed-style interface, while those in the \textit{CommSense} condition were asked in greater detail about the roles of specific features in shaping their judgment, as well as suggestions for improvement.

\subsection{User Study Results}
\par We collected participants' ratings on the three main dimensions outlined in \cref{sec:evadims}: \textit{Perception}, \textit{Functionalities}, and \textit{Judgment Assessment}. All ratings were measured on a 7-point Likert scale. Quantitative data for \textit{Perception} and \textit{Judgment Assessment} were analyzed using the Mann–Whitney U Test~\cite{mcknight2010mann} to compare \textit{CommSense} with the baseline interface. Ratings of \textit{Functionalities} were summarized descriptively without formal statistical testing. For the qualitative data, interview transcripts were analyzed using inductive thematic analysis~\cite{braun2012thematic}, with two researchers cross-checking the coding for consistency. This mixed-methods design enabled us to capture key themes and patterns in participants' feedback, thereby complementing and contextualizing the quantitative findings.

\begin{table*}[h]
\centering
\caption{User ratings on \textit{CommSense} functionalities.}
\label{tab:commense_functionalities}
\begin{tabular}{lcccc} 
\hline
\textbf{Functionalities~} & \textbf{DR} & \textbf{Descriptions} & \textbf{Mean} & \textbf{Std} \\ 
\hline
Overview Providing & \textbf{DR1} & \begin{tabular}[c]{@{}c@{}}Provide an overview of comments and key patterns,\\~form balanced initial impressions.\end{tabular} & 6.25 & 0.62 \\ 
\hline
Comment Organizing & \textbf{DR2} & ~Support organizing comments & 6.08 & 0.67 \\
Evidence Collecting & \textbf{DR2} & Support annotating comments snippets & 6.33 & 0.78 \\
Reminder Triggering & \textbf{DR3} & \begin{tabular}[c]{@{}c@{}}Help users identify inconsistencies or \\recurring themes in the workspace.\end{tabular} & 6.08 & 0.79 \\ 
\hline
Judgment Synthesizing & \textbf{DR4} & \begin{tabular}[c]{@{}c@{}}Support presenting users’ evolving assessments \\and synthesizing their judgments.\end{tabular} & 6.17 & 0.58 \\
\hline
\end{tabular}
\end{table*}

\subsubsection{Perception of \textit{CommSense}}
\par The perception of \textit{CommSense} was assessed across several dimensions, including system usability (SUS) and cognitive workload (NASA-TLX), with results benchmarked against a baseline interface resembling standard online comment sections. Statistical analyses were performed with a significance threshold of $p < 0.05$. \cref{tab:commense_perception} reports the detailed results, where * indicates $p < 0.05$ and ** indicates $p < 0.01$, along with means ($M$) and standard deviations (SD) for each metric. Overall, \textit{CommSense} showed significant improvements over the baseline in facilitating structured comment navigation, reducing cognitive load, and enhancing user engagement. The full set of survey questions is provided in \cref{sec:perc_comm}.

\par \textbf{System Usability.} System usability of \textit{CommSense} was evaluated across multiple factors, including ease of use, functionality, learnability, frequency of use, user confidence, and perceived inconsistency. As shown in \cref{tab:commense_perception}, participants rated \textbf{functionality} significantly higher for \textit{CommSense} ($M = 6.17$, $SD = 0.72$) compared to the baseline system ($M = 4.50$, $SD = 1.24$, $p = 0.0014$), indicating that its integrated features were viewed as more comprehensive and effective. Similarly, \textbf{frequency of use} was rated higher for \textit{CommSense} ($M = 5.50$, $SD = 1.00$) than for the baseline ($M = 3.67$, $SD = 1.15$, $p = 0.0015$), suggesting stronger user willingness for repeated engagement.

\par Other usability measures, including \textbf{ease of use} ($M = 5.50$, $SD = 1.09$ vs. $M = 5.00$, $SD = 1.41$), \textbf{learnability} ($M = 6.00$, $SD = 0.74$ vs. $M = 5.83$, $SD = 0.94$), \textbf{confidence} ($M = 6.17$, $SD = 0.83$ vs. $M = 5.17$, $SD = 1.34$), and \textbf{perceived inconsistency} ($M = 2.17$, $SD = 1.34$ vs. $M = 2.75$, $SD = 1.22$), also showed favorable trends for \textit{CommSense}, although these differences were not statistically significant.

\par \textbf{Cognitive Workload.} Cognitive workload was evaluated using the NASA-TLX~\cite{hart1988development} across six dimensions: mental demand, physical demand, temporal demand, performance, effort, and frustration. As summarized in \cref{tab:commense_perception}, \textit{CommSense} significantly reduced \textbf{mental demand} ($M = 3.17$, $SD = 1.27$ vs. $M = 4.58$, $SD = 1.00$, $p = 0.008$), \textbf{physical demand} ($M = 2.00$, $SD = 0.74$ vs. $M = 2.92$, $SD = 1.16$, $p = 0.041$), and \textbf{temporal demand} ($M = 2.58$, $SD = 1.08$ vs. $M = 3.50$, $SD = 0.90$, $p = 0.046$), suggesting that participants experienced lower cognitive load when navigating comments, filtering information, and synthesizing insights. Ratings for \textbf{performance} ($M = 6.00$, $SD = 0.95$ vs. $M = 5.17$, $SD = 1.27$), \textbf{effort} ($M = 2.58$, $SD = 1.08$ vs. $M = 3.50$, $SD = 0.90$), and \textbf{frustration} ($M = 1.58$, $SD = 0.51$ vs. $M = 1.83$, $SD = 0.58$) did not differ significantly. Overall, these findings indicate that \textit{CommSense} effectively reduces cognitive workload while maintaining a functionally rich interface that supports reflective and bias-aware judgment of online comments.

% Taken together, these results indicate that CommSense effectively balances usability and cognitive efficiency: it reduces mental, physical, and temporal demands while providing a comprehensive, functionally rich interface that supports reflective and bias-aware evaluation of online comments. Participants reported greater willingness to engage repeatedly with the system, improved confidence, and smoother navigation, demonstrating that CommSense enhances structured exploration and the overall quality of user interaction with online comment data.

% \vspace{-1mm}
\subsubsection{Functionalities of \textit{CommSense}}
\par The functionality of \textit{CommSense} was evaluated across five key aspects: overview providing, comment organizing, evidence collecting, reminder triggering, and assessment synthesizing. Each aspect corresponds to the design requirements (\textbf{DRs}) that informed the system's development and reflects how participants engaged with specific capabilities throughout the tasks. Unlike the perception measures, functionalities were assessed descriptively, emphasizing user ratings and qualitative feedback rather than direct comparisons with the baseline. The results are summarized in \cref{tab:commense_functionalities}. The full set of survey questions is provided in \cref{sec:Functionalitiesofco}.

\par \textit{Functionality Assessment: Overview Providing.} The \textit{Overview Providing} functionality in the \textit{Topic Corpus Overview}, corresponding to \textbf{DR1}, received a high average score ($M = 6.25$, $SD = 0.62$), indicating that participants appreciated the structured summary of comments and key patterns. Many noted that the overview facilitated balanced initial impressions by quickly highlighting both positive and negative perspectives. For example, P7 stated, ``\textit{The overview gave me a clear picture of what people were saying as a whole, so I felt I wasn't missing the broader context.}'' Similarly, P4 remarked, ``\textit{I noticed the comment I was focusing on was just a small part of the whole context from the overview, so it couldn't represent most of the comments.}'' A few participants (e.g., P2, P9) mentioned that the overview could sometimes feel cognitively demanding and suggested ways to streamline exploration. Participants also appreciated the progressive masking of hexagons corresponding to viewed comments, which provided immediate visual feedback on explored areas and encouraged further engagement with the content. Overall, \textit{Overview Providing} was perceived as an effective starting point for forming comprehensive impressions.

\begin{table*}[h]
% \vspace{-1.6mm}
\caption{Time spent by users on different components of the two systems. This table provides a breakdown of user interaction time, including the total session duration and the time (in seconds) dedicated to each system component, for both the baseline and \textit{CommSense}. Values are presented as Mean/Standard Deviation (S.D.).}
\vspace{-2.2mm}
\normalsize
\renewcommand{\arraystretch}{1.08}
\begin{tabular}{cccc}
 \hline
\begin{tabular}[c]{@{}c@{}}\textbf{System/}\\ \textbf{Component}\end{tabular} & \textbf{Metric} & \begin{tabular}[c]{@{}c@{}}\textbf{Baseline}\\ \textbf{Mean/S.D.}\end{tabular} & \begin{tabular}[c]{@{}c@{}}\textbf{\textit{CommSense}}\\ \textbf{Mean/S.D.}\end{tabular} \\
\hline
Total Time & Time (s) & 1143.23 / 721.24 & 1261.77 / 405.61 \\
\hline
\multirow{2}{*}{Comment Navigation Panel} & Time (s) & 1012.01 / 632.82 & 997.74 / 429.64 \\
 & Percentage (\%) & 88.60 / 3.97 & 76.83 / 17.00 \\
\hline
\multirow{2}{*}{Topic Corpus Overview} & Time (s) & 56.34 / 48.51 & 114.02 / 77.46 \\
 & Percentage (\%) & 4.33 / 3.28 & 10.35 / 8.57 \\
\hline
\multirow{2}{*}{Synthesis Board} & Time (s) & 74.89 / 74.03 & 150.01 / 106.06 \\
 & Percentage (\%) & 7.07 / 5.56 & 12.81 / 11.03\\

\hline
\end{tabular}
\label{tab:componentTime}
\end{table*}

\par \textit{Functionality Assessment: Comment Organizing and Evidence Collecting.} The comment organizing and evidence-collecting functionalities, aligned with \textbf{DR2}, support users in managing and annotating individual comments. The comment organizing feature received an average rating of $M = 6.08$ ($SD = 0.67$), while evidence collecting scored $M = 6.33$ ($SD = 0.78$). Participants consistently highlighted the utility of clustering comments by themes and marking important excerpts. P2 noted, ``\textit{Organizing comments into groups made my reasoning much clearer. For example, some people said the hotel's free wine was great, while others mentioned it was only available for a limited time ... Seeing all of this in the group made it really easy for me to notice how different people felt about the same details and spot patterns across viewpoints.}'' Similarly, P3 remarked, ``\textit{I used to just focus on the bad comments, but now I see that the good ones have useful information too. They offset the negatives a bit, and my perspective feels much more balanced.}'' Some participants (e.g., P10, P11) suggested collapsing redundant comments so they appear only once, which could reduce workload and facilitate information processing. Meanwhile, several participants (P1, P4, P5) found \textit{Evidence Collecting} valuable for grounding their judgments. P5 explained, ``\textit{It helps me capture my thoughts, leave a record I can refer back to, and easily note the parts I find important while navigating.}'' Overall, these functionalities were perceived as critical for structured reasoning and effective engagement with the comment corpus.

\par \textit{Functionality Assessment: Reminder Triggering.} The reminder triggering feature, aligned with \textbf{DR3}, received an average rating of $M = 6.08$ ($SD = 0.79$). Participants valued its ability to surface inconsistencies and highlight recurring themes during their analysis. P6 explained, ``\textit{The reminders nudged me to reconsider when I overlooked contradictory comments.}'' Similarly, P10 remarked, ``\textit{It was like having a second pair of eyes catching what I missed.}'' P1 added, ``\textit{It appears very naturally, and I'm willing to keep reading. It provides objective statistics and an AI summary, which basically matches my impressions. I'm willing to take its suggestions and consider other perspectives.}'' Some participants (e.g., P9, P12) suggested that the experience could be further refined through more precise and flexible customization. P9 noted, ``\textit{I generally understood the purpose of reminders, but personal priorities could differ for certain topics.}'' For example, although most comments about pet-friendly hotels tend to be positive, P9 was especially concerned with ``\textit{nighttime noise, such as dogs barking.}'' In such cases, reminders emphasizing general sentiment or common patterns could feel slightly intrusive, though participants still found them useful. Overall, the reminder functionality effectively encouraged reflective engagement and supported participants in forming more balanced judgments.

\par \textit{Functionality Assessment: Judgment Synthesizing.} The judgment synthesizing feature, aligned with \textbf{DR4}, achieved an average rating of $M = 6.17$ ($SD = 0.58$). This functionality allowed participants to consolidate their evolving assessments into a coherent final judgment. Many participants emphasized its value in structuring and clarifying their reasoning. P8 noted, ``\textit{The synthesis board helped me tie everything together; it felt like a natural place to finalize my answer.}'' Similarly, P1 remarked, ``\textit{It kept me from jumping to conclusions too early by reminding me to reflect on all the evidence.}'' A few participants (e.g., P3, P4) suggested enhancements to move beyond simple note-taking. They proposed mechanisms for systematically organizing information and tracking the evolution of their reasoning over time, such as a tree-like structure that visually maps how their judgments develop across different pieces of evidence. These improvements could further support reflective and structured decision-making.

% \begin{wrapfigure}{r}{0.5\textwidth}
% \centering
% \begin{tabular}{cc}
% \includegraphics[width=0.45\linewidth]{figures/baselineSemanticHeatmap.png} &
% \includegraphics[width=0.45\linewidth]{figures/commsenseSemanticHeatmap.png}\\
% (a) Baseline & (b) CommSense\\
%  \multicolumn{2}{c}{  \includegraphics[width=0.95\linewidth]{figures/PDF.png}} \\
%  \multicolumn{2}{c}{(c) } \\
%  \end{tabular}
% \caption{}
% \label{fig:semanticHeatmap}
% \label{fig:PDF}
% \end{wrapfigure}

% \begin{wrapfigure}{r}{0.5\textwidth}% \vspace{-4mm}
%     \centering
%     \includegraphics[width=0.5\textwidth]{figures/PDF.png}
%     \caption{PDF}
%     
%     \vspace{-6mm}
% \end{wrapfigure}

\par Besides the specific \textit{Functionality Assessment} conducted on \textit{CommSense}, \textit{Interaction Behaviors} were analyzed based on participants' usage logs across both systems in terms of \textit{usage time across panels} and \textit{comment coverage}.

\begin{figure}
  \centering
  \includegraphics[width=\columnwidth]{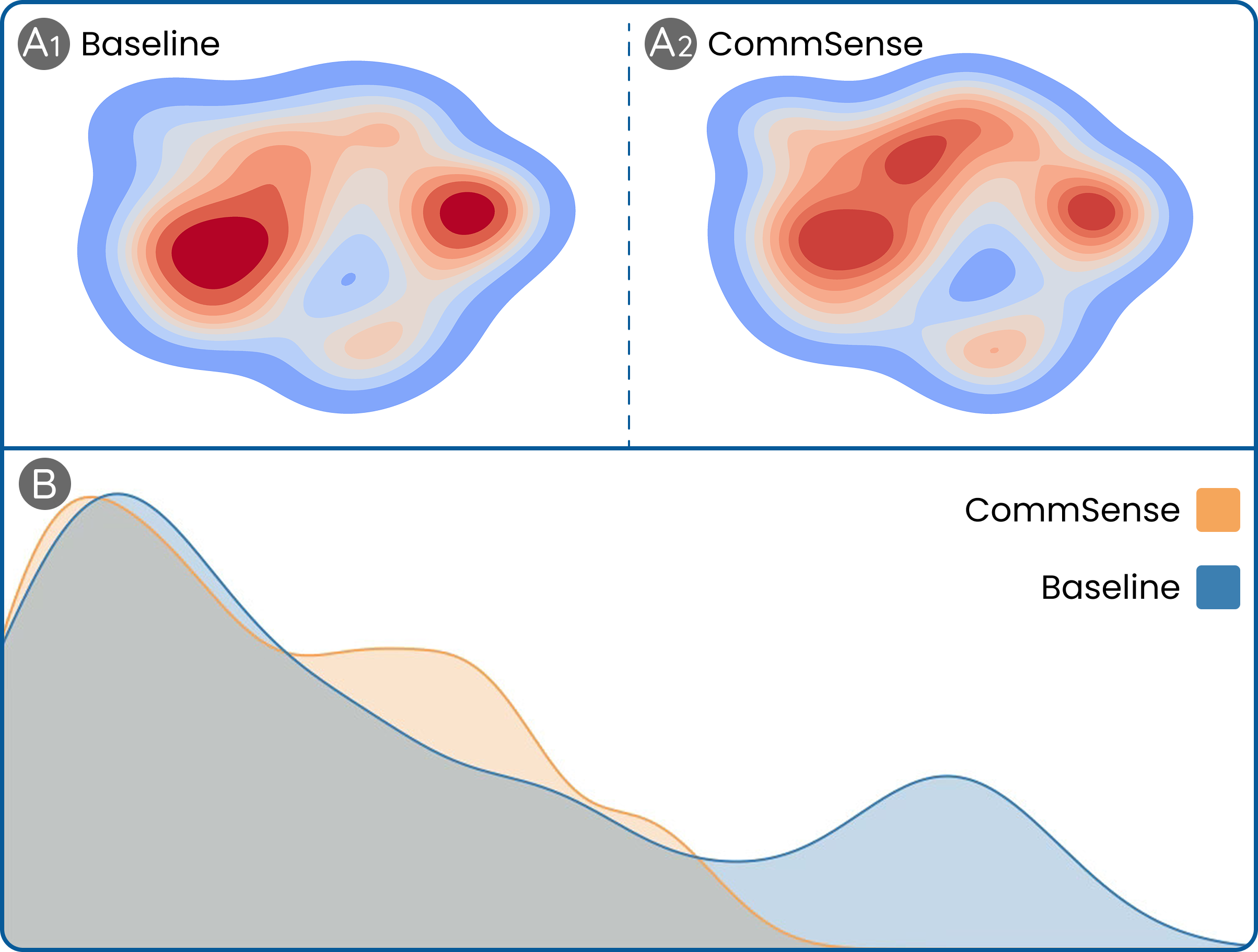}
    \caption{Visualizations of user reading behavior on comments. (A) Heatmap showing the semantic distribution of read comments. The color gradient indicates semantic weight, with blue corresponding to low weights and red to high weights. A1 is the baseline condition, and A2 is for \textit{CommSense}. (B) The distribution of comment read frequency (smoothed by Kernel Density Estimation). The x-axis represents the comment read frequency, and the y-axis represents the density. CommSense is represented in orange, while the baseline is in blue.}
  \label{fig:logAnalysis}
\end{figure}

\par \textit{Usage time across panels.} As shown in \cref{tab:componentTime}, participants spent slightly more time using \textit{CommSense} than the baseline, largely due to greater engagement with the \textit{Overview} and \textit{Synthesis} boards, where usage was nearly twice as high. This result highlights the effectiveness of the two panels and is consistent with the user feedback reported above.

\par \textit{Comment Coverage: Semantic and Sentiment Distribution.} We further analyzed the sentiment and semantic distribution of comments accessed by participants. For sentiment (\cref{sec:sentiment_analysis}), the Positive-to-Negative Ratio with \textit{CommSense} ($M=1.44$, $SD=0.24$) was closely aligned with the original dataset ($M=1.47$), and significantly higher than that of the baseline ($M=1.13$, $SD=0.25$, $p=0.0133$)\footnote{Statistical significance was determined based on the absolute difference between each condition (\textit{CommSense} or baseline) and the original dataset.}. For semantics (see \cref{fig:logAnalysis}), participants using \textit{CommSense} engaged with a broader and more balanced set of comments. Specifically, in the semantic space (\cref{fig:logAnalysis}-A1 vs. \cref{fig:logAnalysis}-A2), baseline interactions were concentrated within a few clusters, whereas those with \textit{CommSense} were distributed across a wider range of regions, indicating broader topic coverage. Complementing this, the frequency distribution in \cref{fig:logAnalysis}-B shows that baseline usage was skewed toward a limited subset of comments, while \textit{CommSense} produced a smoother distribution, suggesting a more equal likelihood of each comment to be read.

%%% 
% 视图使用时间分布

% 语义分布

% \begin{wrapfigure}{r}{0.5\textwidth} 
%     \vspace{-6mm}
%     \centering
%     \includegraphics[width=0.52\textwidth]{figures/quality_of_user_study.png}
%     \caption{Final Judgment Assessment: comparing the baseline and \textit{CommSense} across Logical Consistency, Completeness, Reflective Depth, and Evidence Support.}
%     \label{fig:placeholder}
%     \vspace{-6mm}
% \end{wrapfigure}

\begin{figure}[h]
    % \vspace{-6mm}
    \centering
    \includegraphics[width=\columnwidth]{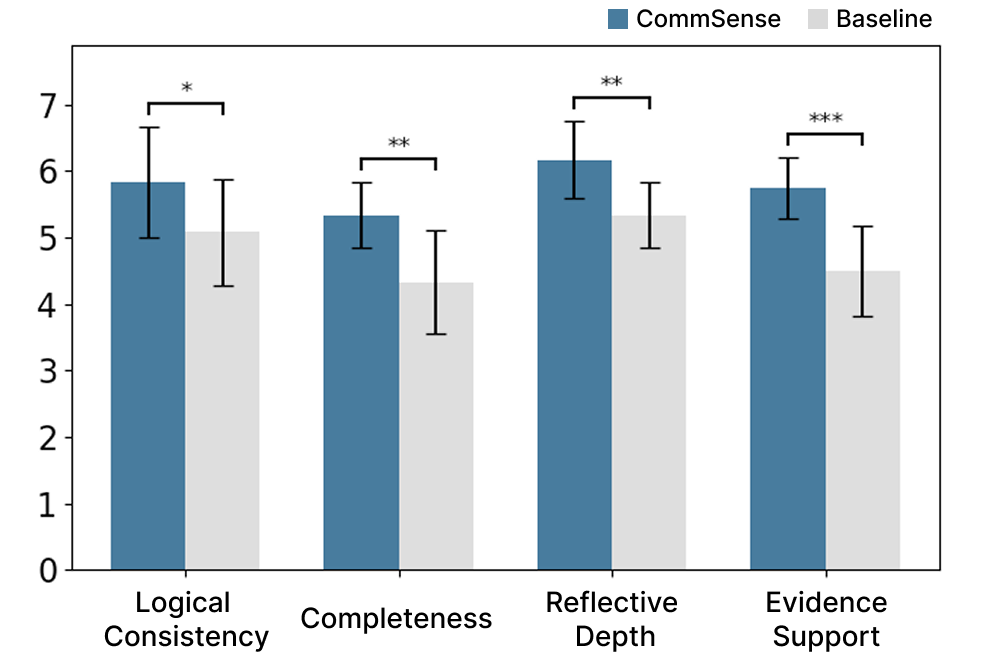}
    \caption{Final Judgment Assessment: comparing the baseline and \textit{CommSense} across Logical Consistency, Completeness, Reflective Depth, and Evidence Support.}
    \label{fig:jud_ass}
    \vspace{-2mm}
\end{figure}

\subsubsection{Final Judgment Assessment}
\par Detailed questions are provided in \cref{sec:Judgmentsofcomm}. In terms of \textit{Final Judgment Assessment} (see \cref{fig:jud_ass}), participants rated their peers' final judgments more favorably across all four dimensions—Logical Consistency, Completeness, Reflective Depth, and Evidence Support—when using \textit{CommSense}. \textit{CommSense} consistently achieved higher scores across all dimensions, with an average score of 5.77 compared to 4.81 for the baseline. Notably, in the \textit{Reflective Depth} category, \textit{CommSense} showed substantial improvement, with an average score of 6.17 compared to 5.33 for the baseline. These results indicate that \textit{CommSense} supported the creation of higher-quality, more reflective, and evidence-grounded judgments.

% \subsubsection{}

% 重复的酒店 评论 怎么并在一起
% 无侵入式 看总体分布
% 探索流数据状态【不断有新的】长效的情况 debias

% 7.1 揭示偏见感知的内在结构：从静态现象到动态过程
% 以往研究多将认知偏见视为一个静态的决策结果，而我们的研究则揭示了其动态的形成过程 (a dynamic process)。研究I中发现的**“四阶段决策路径”**（初始建构、证据搜寻、信念更新、综合判断）是本次讨论的基石。它提供了一个分析框架，清晰地展示了“锚定效应”等偏见并非一次性发生，而是在“初始建 usada”阶段就已播下种子，并持续扭曲后续的整个信息处理流程。

% 我们发现，早期的评论不仅设定了用户的情感基调，更关键的是，它“异化”了用户的认知目标——从开放的“探索”转变为封闭的“验证”或“防御”。这一根本性的转变，解释了为何用户会陷入**“认知过度简化”（过早下结论）或“认知超载”**（持续犹豫不决）的困境。更重要的是，这个模型指出了现有界面在“综合判断”阶段的普遍缺失，正是这种支持的缺位，使得用户在决策的“最后一公里”更容易退回到启发式思维。因此，本研究的核心贡献之一，便是将偏见问题从“是什么”的层面，推进到了“如何发生”的层面，为设计干预提供了清晰的路线图。

% 针对 RQ1，我们发现，朴素的评论界面容易放大用户对突出或极端评论的依赖，导致bias。我们讲用的认知轨迹归纳decision path，分成几个关键的阶段。其中各个阶段有其自己的challengg,带来bias的风险。算法自身的排序会使得部分评论被过度关注，而其他同样重要的评论证据被忽视会被忽视。这与既有研究一致 [xx, xx]，表明界面设计会显著影响用户在决策中的信息取舍与偏差水平。相比之下，我们的tool-CommSense 通过更均衡的评论组织与概览呈现，帮助用户在多条证据之间保持全面视角，从而减少仅凭显著或情绪化评论做出判断的倾向。
% \newpage
\section{Discussion and Limitation}
% \subsection{Revealing the Internal Dynamics of Bias Perception in Comment Navigation}
% Our findings suggest that the cognitive trajectory in comment navigation unfolds as a dynamic, multi-stage process rather than a static outcome. In baseline interfaces resembling standard comment sections, participants’ initial impressions were strongly shaped by salient or extreme comments, which subsequently guided selective evidence seeking and biased belief updating—patterns that align closely with the four-stage decision path identified in Study I.
% \subsection{Serve as an On-the-fly Plugin}
\subsection{Serve as an On-the-fly Plugin}
% 跟comment analysis的区别
\par Participants in the \textit{CommSense} group reported that the plugin seamlessly integrated into their browsing workflows, providing real-time access to aggregated overviews and gentle reminders. However, its benefits may vary across different comment-reading scenarios, as the plugin is intended to aid users in forming their own critical judgments rather than replace their evaluation process~\cite{wu2025comviewer}. The in-situ assistance helped users identify contradictory opinions and recurring patterns, promoting more balanced and evidence-based judgments rather than serving as a tool for casual exploration or entertainment~\cite{cai2024pandalens}.

\par Unlike traditional data analysis tools that focus primarily on extracting, aggregating, and visualizing comment statistics and insights~\cite{yan2024commentvis}, \textit{CommSense} emphasizes user-centered, in-situ support. 
% Its lightweight and unobtrusive design ensures minimal disruption to the primary task of reading and interpreting comments. By externalizing memory and reasoning processes, the plugin allows users to concentrate on evaluating the relevance and usefulness of information rather than managing the cognitive load of information processing.
\ouyangre{Its lightweight and unobtrusive interface reduces cognitive load while scaffolding reflective evaluation~\cite{menon2020nudge}, serving as a concrete example of how HCI interfaces can externalize reasoning and foster bias-aware sensemaking in unstructured, text-heavy environments.}

\par Several participants (P2, P4, P6, P10) also described \textit{CommSense} as engaging and interesting, reflecting patterns of sustained use and long-term engagement. 
% These findings suggest that \textit{CommSense} not only supports reflective evaluation but also encourages continued exploration of comments~\cite{tan2025curious}.
\ouyangre{These observations suggest that \textit{CommSense} not only facilitates on-the-fly reflective evaluation but may also cultivate habitual strategies for critically assessing user-generated content, highlighting the potential for interactive systems to shape both momentary judgments and longer-term information evaluation behaviors~\cite{tan2025curious}.}

\subsection{Empower Users as Rational and Reflective Sensemakers}
\subsubsection{Enhancing Bias Awareness and Critical Thinking}
\par Participants valued the aggregated overview provided by \textit{CommSense}'s \textit{Topic Corpus Overview}, noting that it allowed them to quickly understand the distribution of opinions, major clusters, and outlier perspectives before engaging with individual comments. This feature helped participants identify potential biases in their initial impressions, supporting more critical and reflective evaluation of the comment corpus. These findings are consistent with prior work suggesting that aggregated visualizations can aid users in recognizing and correcting cognitive biases in information-rich environments~\cite{ellis2018cognitive}. Participants also appreciated the \textit{In-situ Reminders}, which delivered subtle, context-sensitive nudges~\cite{caraban201923} when the system detected patterns of selective reading or rapid judgment. For example, when users focused primarily on extreme positive or negative comments, the reminders encouraged them to consider underrepresented viewpoints. This aligns with previous research demonstrating that subtle, context-aware interventions can foster more balanced information processing~\cite{rieger2024nudges}.

\subsubsection{Alleviating Cognitive Load During Complex Exploration}
\par A common concern with tools designed to encourage deeper thinking is that they may increase users' operational and cognitive load~\cite{gerlich2025ai}. However, results from Study III suggest that \textit{CommSense}, while guiding users toward more complex thought processes, does not substantially increase, and may even slightly reduce, mental, physical, and temporal demands. This is achieved through features such as \textit{Comment Organization} and \textit{Evidence Snippets}, which decompose the complex task of meaning-making into lightweight, visual operations. By externalizing memory and reasoning processes, \textit{CommSense} alleviates the cognitive burden of ``remembering information'', allowing users to focus on more strategic and complex aspects of their task, thereby making deep thinking more efficient~\cite{skulmowski2023cognitive}.

\subsubsection{Supporting Evidence-Based Argument Construction}
\par Participants found the \textit{Synthesis Board} particularly valuable for consolidating spontaneous \textit{Instant Thoughts}, highlighted Evidence Snippets, and system-generated Reminders into a unified workspace. They reported that it helped them visualize connections, identify contradictions, and iteratively refine their reasoning, supporting the construction of coherent and well-supported judgments. Peer-assessment results from Study III further indicated that judgments produced using the \textit{Synthesis Board} were more logically consistent, better supported, and more robust than those created with conventional tools. This supports prior research demonstrating that structured, integrative panels can scaffold complex reasoning and improve judgment quality~\cite{ayal2015determinants}.
\par Nevertheless, participants expressed a desire for greater customization. For instance, they preferred the ability to adjust how evidence and reminders are displayed, filter or prioritize certain types of inputs, and modify the workspace structure to better suit their reasoning strategies. Such feedback underscores the importance of adaptive and user-tailored interfaces in enhancing engagement and supporting diverse cognitive processes~\cite{todi2021adapting}.

\subsection{Integrating LLMs in \textit{CommSense}}
\par In our tool, LLMs (see \cref{sec:imdetails}) are used to perform topic summarization in the \textit{Topic Corpus Overview} and to generate AI summaries for the \textit{Reminder Module}, providing concise insights and context-aware prompts that support users' reflective evaluation of comments. Despite their utility, current LLMs have inherent limitations, including occasional inconsistencies across iterations and the risk of generating inaccurate or fabricated information (hallucinations)~\cite{huang2025survey,ji2023towards}. To mitigate these risks, the \textit{Reminder Module} operates in conjunction with statistical grounding, a rule-based approach. By anchoring AI-generated summaries in verifiable evidence~\cite{phillips2021four}, the design enables users to critically evaluate validity while retaining control over their judgments—a feature that participants in our study generally appreciated and trusted.

\par Additionally, two participants explicitly expressed a desire for summaries and suggestions tailored to their predefined evaluation preferences. This highlights the need for more personalized options, such as fine-tuning the model with user-provided materials to better address individual needs~\cite{hu2023llm}. Looking forward, a more proactive and personalized chatbot~\cite{jones2024designing,ma2024beyond} could provide a companion-like reading experience by actively generating summaries, highlighting key information, and prompting reflection, while still posing challenges in maintaining transparency and user control~\cite{hasal2021chatbots}.

\subsection{Generalizability}
% We used hotel comments as a controlled yet realistic proxy to evaluate interaction strategies. Unlike real hotel-booking scenarios, which often include explicit ratings and structured comparisons, our study deliberately omitted these signals to observe how participants navigate unstructured, text-heavy comments. This design choice may, in fact, enhance the generalizability of our findings: the four-stage decision path observed in Study I highlights recurring challenges in navigating unstructured information, suggesting that these patterns extend beyond the specific setting studied~\cite{wu2025comviewer,cai2024pandalens}.
\par We used hotel comments as a controlled yet realistic proxy to evaluate interaction strategies. Unlike real hotel-booking scenarios, which often include explicit ratings and structured comparisons, our study deliberately omitted these signals to observe how participants navigate unstructured, text-heavy comments. This design choice does not aim to replicate real-world conditions exactly but may, in fact, enhance the generalizability of our findings. In fact, the four-stage decision path observed in Study I highlights recurring challenges in navigating unstructured information, suggesting that these patterns extend beyond the specific setting studied.

% \par The design strategy of \textit{CommSense}, with the \textit{Overview} highlighting diverse perspectives, \textit{Reminders} encouraging balanced consideration, and the \textit{Synthesis Board} externalizing reasoning—can be applied to other domains. Tasks such as evaluating health discussions, political debates, or product reviews require integrating large volumes of often conflicting opinions. Its modular, lightweight architecture allows adaptation through algorithmic or visualization adjustments (e.g., credibility cues for health~\cite{sbaffi2017trust,song2019role}, ideological clustering for politics~\cite{fleishman1986types}), supporting more balanced and bias-aware sensemaking across diverse contexts.

The design strategy of \textit{CommSense}—with the \textit{Overview} highlighting diverse perspectives, \textit{Reminders} encouraging balanced consideration, and the \textit{Synthesis Board} externalizing reasoning—embodies cognitive scaffolding principles that guide attention, reduce cognitive load, and support reflective evaluation. From an HCI perspective, these design features illustrate how lightweight, in-situ assistance can be integrated into user workflows to promote bias-aware sensemaking, a principle that can inform other interactive systems~\cite{yan2024commentvis,tan2025curious}. \ouyangre{Such strategies can be adapted to domains requiring the integration of large volumes of often conflicting opinions, including health discussions~\cite{sbaffi2017trust,song2019role}, political debates~\cite{fleishman1986types}, or product reviews. Its modular architecture allows algorithmic or visualization adjustments, such as credibility cues for health~\cite{sbaffi2017trust,song2019role} or ideological clustering for politics~\cite{fleishman1986types}. Scaling such systems, however, introduces challenges: maintaining real-time responsiveness, avoiding user over-reliance, preserving non-intrusive assistance, and accommodating user diversity. Future research could explore adaptive scaffolding that modulates support based on user expertise or detected biases, extending the applicability of these design principles to varied contexts~\cite{wu2025comviewer,cai2024pandalens}.}

\subsection{Limitation}
\par Our work has several limitations. First, our participant pool consisted primarily of graduate students, who may possess higher cognitive abilities and reflective motivation than the general population. Future research should evaluate \textit{CommSense} with more diverse populations across different ages, educational backgrounds, and cultural contexts to assess its broader applicability. 
\ouyangre{Second, using the same participants in both Studies I and II may have influenced these participants’ attention and priorities, potentially affecting the resulting design outcomes. While this practice is common in formative HCI research, it may introduce demand‑characteristic effects or response biases~\cite{nichols2008good,dell2012yours}. Future work should evaluate these design concepts with larger and independent participant samples.} 
Third, while \textit{CommSense}'s \textit{Topic Corpus Overview} effectively summarizes opinion distributions, it may introduce bias by emphasizing dominant clusters or underrepresenting minority viewpoints~\cite{wall2019toward}. Future work could enhance transparency in aggregation methods or explicitly highlight underrepresented perspectives to avoid inadvertently biasing interpretation~\cite{hasanzadeh2025bias}. Finally, our study was limited by its controlled, single-session experimental design. The simplified proxy task may not fully capture real-world complexities, and the long-term effects of prolonged use—such as user dependency or diminished reminder effectiveness—remain unknown. Future longitudinal studies~\cite{caruana2015longitudinal} should therefore be conducted in more general and authentic scenarios to fully assess the real-world impact and applicability of \textit{CommSense}.
% 尽管本研究取得了积极成果，但仍存在一些局限性，这也为未来的研究指明了方向：
% 1. 参与者群体的泛化性： 我们的研究参与者主要是研究生，他们可能比普通大众拥有更高的认知能力和反思动机 。未来的研究需要在更广泛、更多样化的人群中（例如，不同年龄、教育背景和文化背景的用户）对CommSense或类似工具进行测试，以验证其普适性。
% 2. 任务情境的复杂性： 本研究聚焦于“酒店预订”这一相对明确的决策场景 。对于更复杂、争议性更强或情感更激烈的话题（如公共健康信息、政治新闻评论），用户的行为模式和干预效果可能会有所不同。未来可以探索如何将CommSense的设计原则应用于这些更具挑战性的领域。
% 3. 研究的短期效应： 本研究是在单次实验环境中进行的，评估的是工具的即时效果。我们尚不清楚长期使用CommSense会带来何种影响。用户是否会对其功能产生依赖？提醒功能是否会因重复出现而效力减弱？开展一项纵向研究 (Longitudinal Study) 来观察用户行为和思维模式的长期变化将非常有价值。
 % the study focused on a relatively well-defined scenario—hotel evaluation. Users’ behaviors and the tool’s effectiveness may vary in more complex, controversial, or emotionally charged contexts, such as public health information or political news comments. Extending \textit{CommSense} to these domains may require additional design considerations to accommodate diverse user needs and engagement patterns.
% 1. 场景都是酒店评论场景 两个维度 极性比较特殊 医疗场景
% 2.
\section{Conclusion}
\par This study introduced \textit{CommSense}, a tool aimed at facilitating rational judgment by supporting bias-aware and reflective navigation of online comments through pre-engagement framing, interactive organization, reflective prompts, and synthesis support. Insights gained from our preliminary experiment and co-design workshop led to the identification of a four-stage user decision path, the distillation of four design requirements, and the implementation of \textit{CommSense} as an on-the-fly plugin where users can engage with online comments more reflectively. Results from the between-subjects evaluation confirmed that \textit{CommSense} improves bias awareness and reflective thinking, helping users produce more comprehensive, evidence-based rationales while maintaining high usability. Future work will focus on adapting and evaluating \textit{CommSense}'s core interventions for other critical contexts involving conflicting user-generated content, such as health forums, product comments, and political debates, thereby contributing to the goal of designing for more rational online discourse.

\begin{acks}
We gratefully acknowledge the anonymous reviewers for their insightful feedback. This research was supported by the National Natural Science Foundation of China (No. 62372298), the Shanghai Engineering Research Center of Intelligent Vision and Imaging, the Shanghai Frontiers Science Center of Human-centered Artificial Intelligence (ShangHAI), and the MoE Key Laboratory of Intelligent Perception and Human-Machine Collaboration (KLIP-HuMaCo).
\end{acks}

%TC:ignore
\balance
\bibliographystyle{ACM-Reference-Format}
\bibliography{sample-base}

\onecolumn
\appendix
\section{Participants Information in the Study I and Study II}
\label{sec:ParticipantsinstudyI}

% \usepackage{multirow}
\begin{table}[H]
\caption{Detailed information of the participants in the Study I \& II.}
\resizebox{\textwidth}{!}{%
\begin{tabular}{lcccccc}
\hline
\textbf{Condition} & \textbf{ID} & \textbf{Gender} & \textbf{Age} & \textbf{Major} & \textbf{Ongoing Degree} & \textbf{Design Experience (years)} \\
\hline
\multirow{6}{*}{\textbf{Positive-first}} & P1 & Female & 26 & Human-Computer Interaction & Ph.D. & 5 \\
 & P2 & Male & 24 & Human-Computer Interaction & Ph.D. & 3 \\
 & P3 & Male & 24 & Computer Science & Master & 0 \\
 & P4 & Female & 23 & Design & Master & 3 \\
 & P5 & Male & 20 & Human-Computer Interaction & Undergraduate & 1 \\
 & P6 & Female & 21 & Computer Science & Undergraduate & 0 \\
 \hline
\multirow{6}{*}{\textbf{Negative-first}} & P7 & Male & 23 & Human-Computer Interaction & Ph.D. & 2 \\
 & P8 & Female & 26 & Human-Computer Interaction & Ph.D. & 5 \\
 & P9 & Male & 24 & Computer Science & Master & 0 \\
 & P10 & Female & 23 & Human-Computer Interaction & Master & 3 \\
 & P11 & Male & 20 & Computer Science & Undergraduate & 0 \\
 & P12 & Male & 17 & Human-Computer Interaction & Undergraduate & 0 \\
 \hline
\multirow{6}{*}{\textbf{Interleaved}} & P13 & Female & 28 & Design & Ph.D. &  6 \\
 & P14 & Male & 24 & Human-Computer Interaction & Ph.D. & 4 \\
 & P15 & Female & 25 & Computer Science & Master & 3 \\
 & P16 & Female & 24 & Human-Computer Interaction & Master & 2 \\
 & P17 & Male & 20 & Computer Science & Undergraduate & 0 \\
 & P18 & Male & 19 & Computer Science & Undergraduate & 3\\
\hline
\end{tabular}}
\end{table}

\section{Prompt for Topic Extraction}
\label{sec:props_top_ex}

\par The following is the system prompt:
\begin{lstlisting}
You are a professional hotel review analyst. Your task is to categorize 
the provided hotel review keywords into 6 main, high-level categories. 
Each category should have a concise description and list all the 
original keywords that belong to it. The final result must be in JSON 
format. The JSON structure should contain a list named 'categories', 
where each object in the list includes 'category', 'description', and 'keywords' (a list of all keywords for that category).
\end{lstlisting}
The following is the user prompt:
\begin{lstlisting}
Please categorize the following hotel review keywords into 6 classes, 
and list the corresponding keywords for each category:
Keywords: {keywords_string}
Ensure each category name is high-level, provide a brief description, 
and include a 'keywords' list containing all keywords for that category.
Ecach keyword should appear in only one category. 
Example JSON format:
 [
    {
      "category": "CATEGORY_NAME_1",
      "description": "Brief description of this category.",
      "keywords": ["keyword1", "keyword2", "keyword3"]
    },
    {
      "category": "CATEGORY_NAME_2",
      "description": "Brief description of this category.",
      "keywords": ["keyword4", "keyword5"]
    },
    ...
  ]
\end{lstlisting}

\section{Prompt for AI Summary and Suggestion}
\label{sec:props_AI_summary}
% \par The following is the system prompt:
\par The following is the system prompt
\begin{lstlisting}
You are a helpful AI assistant designed to promote reflective and balanced thinking for users analyzing online reviews. """ Your task is to analyze a user's recent reading activity and the content of the reviews they have engaged with. Based on this analysis, you will generate a concise, neutral summary of the key points from the reviews and provide a gentle, actionable suggestion to help the user consider a broader range of perspectives or topics. The final output must be in JSON format, containing two keys: 'summary' and 'suggestion'. """
\end{lstlisting}

\par The following is the user prompt
\begin{lstlisting}
A user is reviewing hotel comments and our system has detected a potential bias in their reading pattern.

Trigger Reason: {trigger_reason}
Reading Statistics: {statistics}
Recently Engaged Comments:
"""
{comment_texts}
"""


Based on the information above, please generate a concise summary of the provided comments and an actionable suggestion to encourage more balanced exploration. The tone should be supportive and helpful, nudging the user to reflect without being prescriptive.
"""
Example JSON format:
{
"summary": "Guests frequently praise the spacious rooms and comfortable beds, often highlighting the great views from the balcony. However, several reviews also mention that the Wi-Fi signal can be weak and the water pressure in the shower is inconsistent.",
"suggestion": "You've focused on some of the room's technical issues. It might be useful to also see what people have said about the comfort and space of the rooms to get a fuller picture."
"""
}
\end{lstlisting}

\section{Participants Information in Study III}
\label{sec:ParticipantsinstudyIII}
% \usepackage{multirow}
% \begin{table}[H]
% \caption{Detailed information of the participants in the Study III}
% \begin{tabular}{lllll}
% \hline
% \textbf{System} & \textbf{ID} & \textbf{Gender} & \textbf{Age} & \textbf{Ongoing Degree} \\
% \hline
% \multirow{12}{*}{\textbf{\textit{CommSense}}} & P1 & Female & 26 & Ph.D. \\
%  & P2 & Male & 25 & Ph.D. \\
%  & P3 & Male & 25 & Ph.D. \\
%  & P4 & Female & 24 & Master \\
%  & P5 & Male & 23 & Master \\
%  & P6 & Male & 23 & Master \\
%  & P7 & Male & 22 & Master \\
%  & P8 & Female & 22 & Master \\
%  & P9 & Female & 21 & Master \\
%  & P10 & Male & 21 & Undergraduate \\
%  & P11 & Female & 20 & Undergraduate \\
%  & P12 & Male & 20 & Undergraduate \\
%  \hline
% \multirow{12}{*}{\textbf{\textit{Baseline}}} & P13 & Male & 28 & Ph.D. \\
%  & P14 & Male & 26 & Ph.D. \\
%  & P15 & Female & 25 & Ph.D. \\
%  & P16 & Male & 25 & Master \\
%  & P17 & Male & 25 & Master \\
%  & P18 & Male & 23 & Master \\
%  & P19 & Female & 22 & Master \\
%  & P20 & Female & 22 & Master \\
%  & P21 & Male & 23 & Master \\
%  & P22 & Male & 21 & Master \\
%  & P23 & Female & 20 & Undergraduate \\
%  & P24 & Female & 18 & Undergraduate\\
%  \hline
% \end{tabular}
% \end{table}

\begin{table}[H]
\caption{Detailed information of the participants in the Study III.}
\begin{tabular}{cccc|
>{\columncolor[HTML]{DAE8FC}}c 
>{\columncolor[HTML]{DAE8FC}}c 
>{\columncolor[HTML]{DAE8FC}}c 
>{\columncolor[HTML]{DAE8FC}}c }
\hline
\multicolumn{4}{c|}{\textbf{\textit{CommSense}}} & \multicolumn{4}{c}{\cellcolor[HTML]{DAE8FC}\textbf{Baseline}} \\ \hline
\textbf{ID} & \textbf{Gender} & \textbf{Age} & \textbf{Ongoing Degree} & \textbf{ID} & \textbf{Gender} & \textbf{Age} & \textbf{Ongoing Degree} \\  \hline
P1 & Female & 26 & Ph.D. & P13 & Male & 28 & Ph.D. \\
P2 & Male & 25 & Ph.D. & P14 & Male & 26 & Ph.D. \\
P3 & Male & 25 & Ph.D. & P15 & Female & 25 & Ph.D. \\
P4 & Female & 24 & Master & P16 & Male & 25 & Master \\
P5 & Male & 23 & Master & P17 & Female & 25 & Master \\
P6 & Female & 23 & Master & P18 & Male & 23 & Master \\
P7 & Male & 22 & Master & P19 & Female & 22 & Master \\
P8 & Female & 22 & Master & P20 & Female & 22 & Master \\
P9 & Female & 21 & Master & P21 & Male & 23 & Master \\
P10 & Male & 21 & Undergraduate & P22 & Male & 21 & Master \\
P11 & Female & 20 & Undergraduate & P23 & Female & 20 & Undergraduate \\
P12 & Male & 20 & Undergraduate & P24 & Female & 18 & Undergraduate\\
 \hline
\end{tabular}
\end{table}

\section{Sentiment Analysis of Comments Read by Participants in Study III.}
\label{sec:sentiment_analysis}
\begin{table}[H]
\caption{Sentiment analysis of comments read by users. The table shows the distribution of positive, neutral, and negative comments, presenting their total counts and percentages for both baseline and \textit{CommSense}. Values are presented as Mean/Standard Deviation (S.D.).}
\label{tab:commentSentiment}
\begin{tabular}{cclcc}
 \hline 
\multirow{2}{*}{\textbf{Sentiment Category}} & \multirow{2}{*}{\textbf{Metric}} & \multirow{2}{*}{\textbf{Dataset}} & \textbf{Baseline} & \textbf{\textit{CommSense}} \\
 &  &  &\textbf{ Mean/S.D.} & \textbf{Mean/S.D.} \\
 \hline 
Total & Count (\#) & 574 & 182.17 / 115.37 & 199.25 / 79.82 \\
\hline
\multirow{2}{*}{Postitve} & Count (\#) & 265 & 75.67 / 47.52 & 96.67 / 39.35 \\
 & Percentage (\%) & 46.17 & 40.70 / 4.44 & 48.02 / 3.70 \\
\hline
\multirow{2}{*}{Neutral} & Count (\#) & 129 & 41.00 / 26.20 & 37.58 / 19.41 \\
 & Percentage (\%) & 22.47 & 22.80 / 1.94 & 18.09 / 2.76 \\
 \hline 
\multirow{2}{*}{Negative} & Count (\#) & 180 & 65.50 / 42.16 & 65.00 / 22.22 \\
 & Percentage (\%) & 31.36 & 36.50 / 3.92 & 33.88 / 4.63\\
 \hline
 Positive-to-Negative Ratio & - & 1.47 & 1.13 / 0.25 & 1.44 / 0.24\\
  \hline
\end{tabular}
\end{table}

\section{Questions about Perception of \textit{CommSense} and Baseline.}
\label{sec:perc_comm}
% \usepackage{multirow}
% \usepackage[normalem]{ulem}
\begin{table}[H]
\caption{Questions about perception of \textit{CommSense} \& Baseline.}
\resizebox{\textwidth}{!}{%
\begin{tabular}{lll}
\hline
\textbf{Categories} & \textbf{Factors} & \textbf{Questions} \\
\hline
\multirow{6}{*}{\textbf{System Usability Scale}} & Easy to use & I thought the system was easy to use. \\
 & Functions & I found the various functions in this system were well integrated. \\
 & Quick to learn & I think most people would learn to use this system very quickly. \\
 & Frequency & I think I would like to use this system frequently. \\
 & Confidence & I felt very confident using the system. \\
 & Inconsistency & I thought there was too much inconsistency in this system. \\
\hline
\multirow{6}{*}{\textbf{NASA Task Load Index}} & Mental Demand & How mentally demanding was the task? \\
 & Physical Demand & How physically demanding was the task? \\
 & Temporal Demand & How hurried or rushed was the pace of the task? \\
 & Performance & How successful were you in accomplishing what you were asked to do? \\
 & Effort & How hard did you have to work to accomplish your level of performance? \\
 & Frustration & How insecure, discouraged, irritated, stressed, and annoyed were you?\\
\hline
\end{tabular}}
\end{table}

\section{Questions about Functionalities of \textit{CommSense}.}
\label{sec:Functionalitiesofco}
\begin{table}[H]
\centering
\caption{Questions about functionalities of \textit{CommSense}.}
\resizebox{\textwidth}{!}{%
\begin{tabular}{ll} 
\hline
\textbf{\textbf{Functionality}} & \multicolumn{1}{c}{\textbf{Questions}} \\ 
\hline
Overview Providing (\textbf{DR1}) & \begin{tabular}[c]{@{}l@{}}The \textit{\textbf{Topic Corpus Overview}}~provided a comprehensive view of comments \\and revealed salient patterns.\end{tabular} \\ 
\hline
Comment Organizing (\textbf{DR2}) & \begin{tabular}[c]{@{}l@{}}The \textbf{\textit{Comment Navigation Panel}}~enabled me to organize and manage comments \\in a structured way.\end{tabular} \\
Evidence Collecting (\textbf{DR2}) & \begin{tabular}[c]{@{}l@{}}The \textbf{\textit{Comment Navigation Panel}}~supported me in annotating and collecting \\useful evidence snippets.\end{tabular} \\
Reminder Triggering (\textbf{DR3}) & \begin{tabular}[c]{@{}l@{}}The \textbf{\textit{Comment Navigation Panel}}~effectively reminded me of inconsistencies \\or recurring themes.\end{tabular} \\ 
\hline
Assessment Synthesizing (\textbf{DR4}) & \begin{tabular}[c]{@{}l@{}}The \textbf{\textit{Synthesis Board}}~facilitated the presentation of my evolving assessments \\and the synthesis of my judgments.\end{tabular} \\
\hline
\end{tabular}}
\end{table}

\section{Questions about Final Judgments of \textit{CommSense} and Baseline.}
\label{sec:Judgmentsofcomm}
\begin{table}[H]
\centering
\caption{Peer scoring questions for final judgment assessment in \textit{CommSense}.}
\begin{tabular}{ll} 
\hline
\textbf{Factors} & \textbf{Questions} \\ 
\hline
Logical Consistency & \begin{tabular}[c]{@{}l@{}}The final judgment is coherent, shows clear reasoning, \\and is logically consistent.\end{tabular} \\
Completeness & \begin{tabular}[c]{@{}l@{}}The final judgment considers multiple perspectives, includes key information, \\and does not omit relevant points.\end{tabular} \\
Reflective Depth & \begin{tabular}[c]{@{}l@{}}The final judgment shows reflection, engages with contrasting opinions, \\and demonstrates careful consideration.\end{tabular} \\
Evidence Support & \begin{tabular}[c]{@{}l@{}}The final judgment is supported by concrete comment evidence, \\linked to relevant examples, and justified by reasoning.\end{tabular} \\
\hline
\end{tabular}
\end{table}

% \section{Evaluation dimensions }
% \begin{table}[ht]
% \centering
% \caption{Three adopted dimensions, along with their categories and corresponding aspects}
% \label{tab:eval_dimensions}
% \begin{tabular}{|l|l|p{6cm}|p{6cm}|}
% \hline
% \textbf{Dimension} & \textbf{Focus} & \textbf{Metrics / Measures} & \textbf{Description} \\
% \hline
% \textbf{Perception} & Subjective experience & SUS, NASA-TLX & Captures overall usability and cognitive workload, providing insight into participants’ impressions of the system. \\
% \hline
% \textbf{Functionalities} & Interaction with system features & Topic selection interactions (\textit{Topic Corpus Overview}), comment view logs, reminder interactions, ``useful'' comment markings (\textit{Comment Navigation Panel}), notes (\textit{Synthesis Board}) & Evaluates how participants explored, navigated, evaluated, and organized information; reflects practical utility and support for engagement and reflection. \\
% \hline
% \textbf{Judgment Assessment} & Quality of final judgments & Logical consistency, completeness, reflective depth, evidence support & Assesses the coherence, thoroughness, and grounding of participants’ judgments in specific comment snippets and multiple perspectives. \\
% \hline
% \end{tabular}
% \end{table}

% \newpage

\section{Codebook for Study I  }
% \section{Final Codebook of High-Level Themes, Subcategories, and Codes for Study I}
\label{sec:codebook}

\begin{table}[h]
% \vspace{-2mm}
\centering
\caption{Final Codebook for High-Level Themes, Subcategories, and Codes}
\label{tab:complete_framework}
% \small % 使用较小字号以容纳全部 55+ 个条目
\renewcommand{\arraystretch}{1.2} % 增加行高，防止文字挤在一起
\begin{tabular}{p{0.25\linewidth} p{0.22\linewidth} p{0.45\linewidth}}
\toprule
\textbf{Theme} & \textbf{Subcategory} & \textbf{Code} \\ \midrule
\multirow{35}{*}{Decision Path} & Initial Framing & • Early anchoring on first impressions \newline • Positive-first confidence boost \newline • Negative-first risk-awareness initiation \newline • Interleaved conditions inducing uncertainty \newline • Formation of initial expectations \newline • Reliance on salient cues \newline • Quick initial attribute ratings \newline • Rapid focus on key attributes \newline • Early sentiment-driven judgments \\ \cmidrule{2-3}
 & Evidence Foraging & • Selective scanning of supportive comments \newline • Deliberate search for contradictory evidence \newline • Strategic exploration of diverse perspectives \newline • Varying depth of comment examination \newline • Attribute re-prioritization \newline • Iterative reading of comment clusters \newline • Dynamic attention allocation across comments \\ \cmidrule{2-3}
 & Belief Updating & • Re-assessment of earlier judgments \newline • Integration of new evidence into beliefs \newline • Frequent upward/downward rating changes \newline • Re-weighting importance of attributes \newline • Reflection on conflicting evidence \newline • Confidence recalibration \newline • Continuous balancing of multiple perspectives \newline • Evidence-driven vs. affect-driven updating \newline • Adjustment of evaluation trajectory based on late information \\ \cmidrule{2-3}
 & Synthesis \& Judgment & • Consolidation of gathered information \newline • Weighing conflicting evidence \newline • Forming final judgments \newline • Mental effort in summarization \newline • Use of heuristics when overwhelmed \newline • Reliance on extreme comments \newline • Comparing across multiple attributes \newline • Evidence-based vs. sentiment-based justification \newline • Confidence in final ratings \newline • Documentation or note-taking behaviors \\ \midrule
\multirow{20}{*}{\shortstack[l]{Decision-\\Making Issues}} & Goal Alienation & • Active verification in positive-first condition \newline • Risk-defense orientation in negative-first condition \newline • Early fixation on strategies reducing flexibility \newline • Overreliance on early salient comments \newline • Selective focus on confirming or disconfirming evidence \\ \cmidrule{2-3}
 & Asymmetric Evidence--Belief Loops & • Confirmatory information seeking \newline • Premature closure or unbalanced scanning \newline • Iterative loops causing cognitive load \newline • Repetition leading to fatigue \newline • Over-simplification \newline • Shallow vs. deep reasoning imbalance \newline • Individual differences in loops \\ \cmidrule{2-3}
 & Insufficient Synthesis Support & • Forgetting prior information \newline • Repeated scrolling \newline • Reliance on heuristics under cognitive load \newline • Difficulty integrating conflicting evidence \newline • Inefficient synthesis process requiring extra effort \newline • Memory limitations potentially leading to biased conclusions \\ \bottomrule
\end{tabular}
% \vspace{-7mm}
\end{table}

%TC:endignore
\end{document}